\documentclass[11pt]{article}       
\usepackage{amsmath}
\usepackage{amsfonts}
\usepackage{amssymb}
\usepackage{amsthm}
\usepackage{lineno}
\usepackage{bigints}
\usepackage{graphicx}
\usepackage{color}
\usepackage{natbib}
\usepackage{apalike}
\usepackage[T1]{fontenc}
\usepackage{multirow}
\usepackage{booktabs}
\usepackage{bbm}
\usepackage{epstopdf}
\usepackage{placeins}
\usepackage{rotating}
\newtheorem{teo1}{Theorem}
\newtheorem{teo2}[teo1]{Theorem}
\newtheorem{teo3}[teo1]{Theorem}
\newtheorem{teo4}[teo1]{Theorem}
\DeclareMathOperator*{\argmax}{arg\,max}
\makeatletter
\def\@seccntformat#1{\@ifundefined{#1@cntformat}%
	{\csname the#1\endcsname\quad}  
	{\csname #1@cntformat\endcsname}
}
\let\oldappendix\appendix 
\renewcommand\appendix{%
	\oldappendix
	\newcommand{\section@cntformat}{\appendixname~\thesection\quad}
}
\makeatother

\title{Flexible modeling of bimodal distributions via skewed-$t$ mixtures \footnote{The \texttt{R} package \texttt{stMix}, containing the codes that implement the methods developed in this paper, is available at \texttt{https://github.com/marco-bee/stMix}.}
}

\author{
	Marco Bee$^{1}$\\
	{\small Corresponding author; ORCID: 0000-0002-9579-3650}
	\and
	Flavio Santi$^{1}$\\
	{\small ORCID: 0000-0002-2014-1981}
	\\[1ex]
	{\small $^{1}$Department of Economics and Management, University of Trento, Italy}
}

\begin{document}
	
	\maketitle
	
\begin{abstract}
We propose a mixture of location-scale skewed-$t$ distributions to fit bimodal, skewed and heavy-tailed data. In particular, the mixture is based on the skewed-$t$ distribution by \cite{fer98}, so that the model-building procedure can be easily extended to mixtures of other symmetric distributions. After studying the properties of the mixture,
we develop a maximum likelihood estimation approach via the EM algorithm and a likelihood ratio test of the null hypothesis of no skewness in any given component. A simulation-based comparison to a recently proposed mixture of g-and-h distributions suggests that the performance of the proposed model is excellent, in terms of both estimation precision in well-specified setups and modeling capability in mis-specified frameworks. Fitting the model to the Standard \& Poor's 500 distortion allows us to confirm the bimodality of its distribution, with the implication that the US stock market has historically been in bearish or bullish conditions, rather than near its fundamental value.
\end{abstract}
	
	
\section{Introduction}

Finite mixture distributions have a long history in statistics, dating back to the pioneering work by \cite{pea94}. From an applied point of view, they have mostly been aimed at modeling bimodal data, which are quite common in real-data applications, especially in medicine and biology: see \citet[Examples 1.3, 9.1.1, 9.4.3]{flu97} or \citet[Example 5.1]{mcl00} and the references therein. In many cases, the natural distribution for this kind of data is a two-population mixture with normal components, which is flexible enough to accommodate bimodality if the component distributions are approximately symmetric and light-tailed \citep{tit85}. However, the normal mixture model must be ruled out if the population components feature skewness and/or leptokurtosis.

Accordingly, for more complex data, mixtures with various non-normal component distributions \citep[Chap. 5]{mcl00} have been developed in the last few decades: in particular, skewed Student-$t$ mixtures that take into account possibly skewed and heavy-tailed components have been devised by \cite{lin07}, \cite{lin10} and \cite{lee14}.
To address the same issue, another approach has recently been proposed by \cite{zha25}. Given the flexibility of the g-and-h distribution (\citealp{hoaglin1985summarizing}), a very flexible model that can accommodate skewed and fat-tailed data, they suggest a two-population mixture of g-and-h populations. The possible drawback is that the g-and-h distribution lacks a closed-form density, so that estimation is not trivial \citep{bee19,bee21a}. As a consequence, fitting a g-and-h mixture is even more difficult; \cite{zha25} propose a quantile-based method for estimation and model selection.

We propose an alternative distribution, given by a two-population mixture of skewed-$t$ location-scale mixtures. This distribution is not less flexible than the g-and-h mixture, but has the advantage of allowing the investigator to implement maximum likelihood estimation (MLE) by means of the EM algorithm. 
Unlike the mixtures of $t$ distributions of \cite{lin07}, \cite{lin10} and \cite{lee14}, our building block is the skewed-$t$ density proposed by \cite{fer98}. In terms of skewness, the flexibility of the proposed skewed-$t$ mixture is due to the introduction of 
inverse scale factors $\gamma$ and  $1/\gamma$ in the positive and the negative orthant of the original density, governing the skewness of each population \citep[Sect. 2]{fer98}. As for kurtosis, it is related to the number of degrees of freedom ($\nu$) of each component distribution, so that it ranges from 3 (i.e., a normal distribution) to arbitrarily large, as $\nu$ gets small.
Full flexibility is obtained by considering a mixture of location-scale skewed-$t$ densities. Estimation can be performed via the EM-algorithm: even though the M-step for some parameters is not in closed form, the numerical maximization required to estimate the parameters is a rather standard problem, which can be tackled by means of classical optimization methods.

It should be noted that our skewed-$t$ mixture is just one of several possible mixtures, whose components densities would be skewed versions of any other symmetric univariate density: the \citet{fer98} mechanism to transform a symmetric density into a skewed one is indeed completely general. The corresponding maximum likelihood estimation procedure can likely be straightforwardly extended as well.

Besides studying the properties of the skewed-$t$ mixture and implementing a maximum likelihood estimation procedure, this paper aims at developing a model that is not less flexible, but analytically more tractable, than the g-and-h mixture, so that it can be estimated more efficiently. In comparative terms, our approach is also well-suited for classification purposes: this property follows directly from the EM-algorithm, which yields estimated posterior probabilities as a by-product.

The rest of this work is organized as follows. In Section \ref{sec:mod_iss} we describe the building block of our model, i.e. the skewed-$t$ distribution, and the corresponding skewed-$t$ mixture. Section \ref{sec:est} develops an EM algorithm for estimation and a test for the null hypothesis of no skewness in any given component. Sections \ref{sec:sim} and \ref{sec:emp} illustrate the outcomes of simulation experiments and of two real-data analyses, respectively. Section \ref{sec:concl} concludes the paper, whereas the appendix contains the proofs of the results in Section 2.

\section{A skewed-$t$-based model for bimodal data}
\label{sec:mod_iss}
\subsection{The skewed-$t$ distribution}

Our goal is to find an appropriate model for bimodal, possibly skewed and/or heavy-tailed, data. Thus, we prefer to identify a general setup that contains different models possibly good at accommodating the aforementioned features.
\cite{fer98} introduce skewness in a zero-mean symmetric density $\tilde{f}$ via a skewness parameter $\gamma$ as follows: 
\begin{align}
	\label{eq:skd}
	f(x;\gamma,\eta)&=\frac{2}{\gamma+\frac{1}{\gamma}}\left\{\tilde{f}\left(\frac{x}{\gamma};\eta\right){\mathbbm 1}_{[0,\infty)}(x)+\tilde{f}(\gamma x;\eta){\mathbbm 1}_{(-\infty,0)}(x)\right\},
\end{align}
where $\tilde{f}$ is a zero-mean symmetric density, $\gamma$ is a shape parameter related to skewness and $\eta$ is a second shape parameter, typically related to tail-heaviness; see \cite{fer98} for details. Notice that the mechanism used for obtaining (\ref{eq:skd}) can be employed for deriving a skewed version of any other zero-mean symmetric distribution. When $\gamma>1$ ($\gamma<1$), $f$ is positively (negatively) skewed, and when $\gamma=1$, $f=\tilde{f}$.


As for $\tilde{f}$, we mention explicitly two choices, among several that can be considered.
\begin{enumerate}
	\item Case 1: $\tilde{f}=:\tilde{f}_{ST}(x;\nu)$ is the central Student $t$ density with $\nu$ degrees of freedom. With this choice, (\ref{eq:skd}) is explicitly given by \citep{gar11}
	\begin{align}
		\label{eq:sktd}
		&f_{ST}(x;\gamma,\nu)=\frac{2}{\gamma+\frac{1}{\gamma}}\frac{\Gamma\left(\frac{\nu+1}{2}\right)}{\Gamma\left(\frac{\nu}{2}\right)(\pi\nu)^{1/2}}\nonumber \\
		&\times \left[1+\frac{x^2}{\nu}\left\{\frac{1}{\gamma^2}{\mathbbm 1}_{[0,\infty)}x+\gamma^2{\mathbbm 1}_{(-\infty,0]}x\right\}\right]^{-(\nu+1)/2}.
	\end{align}
	\item Case 2: $\tilde{f}=:\tilde{\phi}_{ST}$ is the standard normal density. This choice is appropriate if we assume the data to be skewed, but not heavy-tailed.
\end{enumerate}
The four-parameter (location-scale) version of $\tilde{f}$ is straightforwardly obtained as follows:
\begin{align}
	\tilde{f}(x;\mu,\sigma,\gamma,\nu)&=\frac{1}{\sigma}\tilde{f}_{ST}\left(\frac{x-\mu}{\sigma};\gamma,\nu\right)\quad\text{(Case 1)},\nonumber \\
	\tilde{f}(x;\mu,\sigma,\gamma)&=\frac{1}{\sigma}\tilde{\phi}_{ST}\left(\frac{x-\mu}{\sigma};\gamma\right)\quad\text{(Case 2)}.\nonumber
\end{align}
In the following, we will mostly consider the first case. The corresponding location-scale random variable is given by:
\begin{equation}
	\label{eq:skls}
	\tilde{X}\stackrel{\text{def}}{=}\mu+\sigma\tilde{X}_{ST},
\end{equation}
so that $\text{E}(\tilde{X})=\mu$ and $\text{var}(\tilde{X})=\sigma^2\nu/(\nu-2)$, where $\nu>2$; 
see Section \ref{sec:est} for details. Clearly, when $\nu\to\infty$, Case 1 converges in distribution to Case 2. Moreover, significance tests on $\gamma$ allow one to test for the presence of skewness.

For the moment, we restrict ourselves to the case of $t$ distributions with $\mu$ (location) equal to zero and $\sigma$ (scale) equal to one; well known theory of location-scale families will allow us, in the next sections, to easily extend the results obtained in this section to the general case with $\mu\neq0$ and/or $\sigma\neq1$. In the following, we will term the former setup \textit{standardized case}, even though the resulting skewed-$t$ has variance different from one.  

The generality of this approach, not limited to a specific component distribution, clarifies why, among the many versions of the skewed normal \citep{azz05} and skewed-$t$ distribution \citep{lee13} available in the literature, we have chosen the skewed-$t$ distribution proposed by \cite{fer98}, which is obtained by exploiting (\ref{eq:skd}); see also \cite{gar11}. 

\subsection{Mixtures of skewed-$t$ distributions}
\label{sec:back}

Let's now consider the mixture based on the standardized population densities (\ref{eq:sktd}).
The corresponding two-population mixture $X_{ST}$ has density equal to:
\begin{equation}
\label{eq:dens}
h_{ST}(x;\boldsymbol{\theta})=pf_{1,ST}(x;\gamma_1,\nu_1)+(1-p)f_{2,ST}(x;\gamma_2,\nu_2),\ 
\end{equation}
with $x\in\mathbb{R}$ and $p\in(0,1)$.
In this section we derive the cumulative distribution function (cdf), the quantile function, the mean and the variance of the standardized skewed-$t$ mixture
(\ref{eq:sktd}), since they are essential ingredients to find the corresponding functions and quantities of the mixture. Throughout this section, we assume $X_{ST}\sim f_{ST}(x;\gamma,\nu)$, where $f_{ST}(x;\gamma,\nu)$ is the skewed-$t$ density (\ref{eq:sktd}); to avoid trivialities, we also assume $\gamma\neq 1$. All the proofs are reported in Appendix \ref{sec:app}.


\begin{teo1}
\label{teo:teo1}
The cdf of $X_{ST}$ is given by
\begin{equation}
	\label{eq:stcdf}
	F_{ST}(x;\gamma,\nu)=
	\begin{cases}
		\frac{2}{1+\gamma^2}\tilde{F}_{ST}(x\gamma,\nu), & x<0, \\
		\frac{1}{1+\gamma^2} + \frac{2\gamma}{\gamma+\gamma^{-1}} \left(\tilde{F}_{ST}\left(\frac{x}{\gamma},\nu\right)-\frac{1}{2}\right), & x\ge 0,
	\end{cases},
\end{equation}
$\tilde{F}_{ST}(\cdot,\nu)$ is the cdf of the central $t$ distribution with $\nu$ degrees of freedom.
\end{teo1}

To find the cdf of the location-scale version of the distribution, it is enough to notice that, if $F_X(x)$ is the cdf of $X$, the cdf of $a+bX$ is given by $F_X((x-a)/b)$. Hence, in this case we have:
$$
F(x;\mu,\sigma^2,\gamma,\nu)=
\begin{cases}
\frac{2}{1+\gamma_j^2}\tilde{F}_{ST}(x_{ST}\gamma,\nu), & x_{ST}<0, \\
\frac{1}{1+\gamma^2} + \frac{2\gamma}{\gamma+\gamma^{-1}} \left(\tilde{F}_{ST}\left(\frac{x_{ST}}{\gamma},\nu\right)-\frac{1}{2}\right), & x_{ST}\ge 0,
\end{cases}
$$
where $x_{ST}\stackrel{\text{def}}{=}(x-\mu)/\sigma$. 


\begin{teo2}
\label{teo:teo2}
Given $\alpha\in(0,1)$, the quantile function of $X_{ST}$ is given by
$$
F^{-1}_{ST}(\alpha;\gamma,\nu)=
\begin{cases}
	\frac{1}{\gamma}\tilde{F}_{ST}^{-1}\left(\alpha\frac{\gamma^2+1}{2};\nu\right), & \alpha< \frac{1}{2}, \\
	\gamma\tilde{F}^{-1}_{ST}\left(\frac{1}{2\gamma^2} \{\alpha(\gamma^2+1)-1\}+\frac{1}{2};\nu\right), & \alpha\ge \frac{1}{2}.\nonumber
\end{cases}
$$
\end{teo2}

The non-standardized case is easily addressed by exploiting a well known general result: if $X$ is a random variable with  quantile function $F^{-1}_X(\alpha)$, the quantile function of $a+bX$ is given by $a+bF^{-1}_X(\alpha)$. Hence we have
\begin{align}
&F^{-1}(\alpha;\mu,\sigma,\gamma,\nu)=\nonumber \\
&\begin{cases}
	\mu+\sigma\frac{1}{\gamma}\tilde{F}_{ST}^{-1}\left(\alpha\frac{\gamma^2+1}{2}\right), & \alpha<\frac{1}{2}, \\
	\mu+\sigma\gamma\tilde{F}^{-1}_{ST}\left(\frac{1}{2\gamma^2} \{\alpha(\gamma^2+1)-1\}+\frac{1}{2};\nu\right), & \alpha\ge \frac{1}{2}.\nonumber
\end{cases} \nonumber
\end{align}


Both the expected value and the variance can be computed in closed form.


\begin{teo3}
\label{teo:teo3}
The expectation of $X_{ST}$ is given by:
\begin{equation}
	\label{eq:exp}
	\text{E}(X_{ST})=\xi_{\nu,0,+\infty}\left(\frac{\gamma^2-1}{\gamma}\right),
\end{equation}
where $\xi_{\nu,0,+\infty}$ is given by (\ref{eq:condexp}).
\end{teo3}
In the general (i.e., non-standardized) case, (\ref{eq:exp}) becomes
$$
\textnormal{E}(X)=\mu+\sigma\text{E}(X_{ST})=\mu+\sigma\xi_{\nu,0,+\infty}\left(\frac{\gamma^2-1}{\gamma}\right).
$$


\begin{teo4}
\label{teo:teo4}
The variance is given by
\begin{equation}
	\label{eq:var}
	\textnormal{var}(X_{ST})=\xi_{\nu,0,+\infty}^2\left\{\frac{\gamma^6+1}{(\gamma^2+1)\gamma^2}-\left(\frac{\gamma^2-1}{\gamma}\right)^2\right\},\quad \nu>2.
\end{equation}
\end{teo4}
In the non-standardized case, (\ref{eq:var}) is modified as follows:
$$
\text{var}(X)=\sigma^2\text{var}(X_{ST})=\sigma^2\xi_{\nu,0,+\infty}^2\left\{\frac{\gamma^6+1}{(\gamma^2+1)\gamma^2}-\left(\frac{\gamma^2-1}{\gamma}\right)^2\right\},\quad \nu>2.
$$

\subsection{Mixtures of skewed-\textit{t} distributions}

By exploiting the results about the skewed-$t$ distribution in Section \ref{sec:back}, we now turn our attention to the two-population mixture of skewed-$t$ distributions $X$ with density
\begin{equation}
\label{eq:mixdens}
h(x;\boldsymbol{\theta})=p f_1(x;\mu_1,\sigma^2_1,\gamma_1,\nu_1)+(1-p) f_2(x;\mu_2,\sigma^2_2,\gamma_2,\nu_2),
\end{equation}
where $f_j$, $j=1,2$, is the location-scale version of (\ref{eq:sktd}). 
From general results about finite mixture distributions, it is immediate to obtain the cdf:
$$
H(x;\boldsymbol{\theta})=p F_1(x;\mu_1,\sigma^2_1,\gamma_1,\nu_1)+(1-p) F_2(x;\mu_2,\sigma^2_2,\gamma_2,\nu_2).
$$
On the other hand, as always happens when working with finite mixtures, quantiles can only be found numerically: in particular, given $\alpha\in(0,1)$, one has to use a root-finding procedure to solve for $x$ the following equation:
$$
\alpha-H(x;\boldsymbol{\theta})=0.
$$
Even though no explicit solution is available, this is a well-behaved numerical problem, which is guaranteed to have a unique root for $x$ if the cdf is continuous, and can be solved by means of standard numerical root-finding procedures.

The expected value is trivially given by
$$
\textnormal{E}(X)=p \textnormal{E}(X_1)+(1-p) \textnormal{E}(X_2),
$$
where $X_j$ is the skewed random variable used to model the distribution of the $j$-th population ($j=1,2$). Specifically, we exploit Theorem \ref{teo:teo3} to get
\begin{align}
\textnormal{E}(X)&\stackrel{\textnormal{def}}{=}\mu=p \left\{\mu_1+\sigma_1\xi_{\nu_j,0,+\infty}\left(\frac{\gamma_1^2-1}{\gamma_1}\right)\right\}+\nonumber \\
&+(1-p) \left\{\mu_2+\sigma_2\xi_{\nu_j,0,+\infty}\left(\frac{\gamma_2^2-1}{\gamma_2}\right)\right\}.\nonumber
\end{align}

As for the variance, it can be found via the well known general formula \citep[Sect. 2.8]{flu97}
\begin{equation}
\label{eq:mixvar}
\textnormal{var}(X)=\sum_{j=1}^2p_j\textnormal{var}(X_j)+\sum_{j=1}^2p_j(\textnormal{E}(X_j)-\textnormal{E}(X))^2,
\end{equation}
where $p_1=p$ and $p_2=1-p$. In the skewed-$t$ case, by exploiting Theorem \ref{teo:teo4}, (\ref{eq:mixvar}) is equal to:
\begin{align}
\textnormal{var}(X)&=\sum_{j=1}^2p_j\sigma^2\left\{\xi_{\nu_j,0,+\infty}^2\left[\frac{\gamma^6_j+1}{(\gamma_j^2+1)\gamma_j^2}-\left(\frac{\gamma_j^2-1}{\gamma_j}\right)^2\right]\right\}+\nonumber \\
&+\sum_{j=1}^2p_j\left\{\mu_j+\sigma_j\xi_{\nu_j,0,+\infty}\left(\frac{\gamma_j^2-1}{\gamma_j}\right)- \mu\right\}.\nonumber
\end{align}

\section{Estimation and testing}
\label{sec:est}


We estimate the two-population location-scale skewed-$t$ mixture via the EM algorithm. Let $\boldsymbol{x}=(x_1,\dots,x_n)$ be the so-called observed data, i.e. a random sample from the mixture (\ref{eq:mixdens}).
Moreover, let $\boldsymbol{z}=(z_1,\dots,z_n)$ be the unobserved class labels, with $z_i=1$ if the $i$-th observation belongs to the first skewed-$t$ distribution and $z_i=0$ if it belongs to the second; these are the missing data. Moreover, let $\boldsymbol{\theta}=(p,\mu_1,\sigma_1,\gamma_1,\nu_1,\mu_2,\sigma_2,\gamma_2,\nu_2)'$ be the parameter vector. 

With this notation, the observed and complete log-likelihood functions corresponding to (\ref{eq:mixdens}) are respectively given by:
\begin{align}
\ell(\boldsymbol{\theta};\boldsymbol{x})&=\sum_{i=1}^n\log\{pf_1(x_i;\mu_1,\sigma_1,\gamma_1,\nu_1)+(1-p)f_2(x_i;\mu_2,\sigma_2,\gamma_2,\nu_2)\},\nonumber \\
\ell_c(\boldsymbol{\theta};\boldsymbol{x},\boldsymbol{z})&=\sum_{i=1}^nz_i\log p f_1(x_i;\mu_1,\sigma_1,\gamma_1,\nu_1)+\nonumber \\
&\sum_{i=1}^n(1-z_i)\log(1-p) f_2(x_i;\mu_2,\sigma_2,\gamma_2,\nu_2).\nonumber
\end{align}

The algorithm iteratively maximizes the expectation of the complete log-likelihood function, conditional on the current estimates of the parameters and on the observed data; clearly, this requires the computation of the conditional expectation in the first place. For this reason, at the $t$-th iteration, the algorithm is carried out in of two steps \citep{mcl08}.

\medskip
\noindent\textbf{E-step}. Compute the conditional expectation of $\ell_c(\boldsymbol{\theta})$, given the current value of $\boldsymbol{\theta}$ and the observed sample
$\boldsymbol{x}$:
\begin{equation}
\label{eq:Estep0}
Q(\boldsymbol{\theta};\boldsymbol{\theta}^{(t)})\stackrel{\text{def}}{=}\text{E}_{\boldsymbol{\theta}^{(t)}}\{\ell_c(\boldsymbol{\theta})|\boldsymbol{x}\}.
\end{equation}
\textbf{M-step}. Maximize, with respect to $\boldsymbol{\theta}$, the so-called $Q$-function, i.e. the conditional expectation of $\ell_c(\boldsymbol{\theta})$ provided by the E-step (\ref{eq:Estep0}): 
\begin{equation}
\label{eq:Mstep}
\boldsymbol{\theta}^{(t+1)}=\argmax_{\boldsymbol{\theta}} Q(\boldsymbol{\theta};\boldsymbol{\theta}^{(t)}).
\end{equation}

\medskip
The E- and M-step (\ref{eq:Estep0}) and (\ref{eq:Mstep}) are then iterated until convergence is reached according to some stopping rule.
The algorithm monotonically increases the observed likelihood at each iteration; under regularity conditions (\citealp{wu83}), the sequence $\boldsymbol{\theta}^{(t)}$ converges to a stationary point and the estimators are asymptotically efficient.

In the present setup, the conditional expectation of the complete log-likelihood function is given by
\begin{equation}
\text{E}(\ell_c(\boldsymbol{\theta};\boldsymbol{x})|\boldsymbol{x},\boldsymbol{\theta}^{(t)})=\sum_{i=1}^n\tau_{i1}^{(t)}\log p^{(t)}f_1(x_i;\boldsymbol{\theta}_1^{(t)})+\sum_{i=1}^n\tau_{i2}^{(t)}\log (1-p^{(t)})f_2(x_i;\boldsymbol{\theta}_2^{(t)}),
\label{eq:Qfunc}
\end{equation}
where $\boldsymbol{\theta}_i^{(t)}\stackrel{\text{def}}{=}(\mu_i^{(t)},\sigma_i^{(t)},\gamma_i^{(t)},\nu_i^{(t)})$, $i=1,2$, $\tau_{i2}^{(t)}=1-\tau_{i1}^{(t)}$ and
\begin{equation}
\tau_{i1}^{(t)}=\frac{p^{(t)}f_1(x_i;\boldsymbol{\theta}_1^{(t)})}{p^{(t)}f_1(x_i;\boldsymbol{\theta}_1^{(t)})+(1-p^{(t)})f_2(x_i;\boldsymbol{\theta}_2^{(t)})}.
\label{eq:Estep}
\end{equation}
The E-step of the algorithm therefore reduces to the computation of (\ref{eq:Estep}), which is the posterior probability, at the $t$-th iteration, that the $i$-th observation belongs to the first population. As an aside, we note that, at convergence, this quantity can be used for classification purposes.

As for the M-step, it results from the maximization of (\ref{eq:Qfunc}) with respect to all parameters. The M-step for $p$ is given in closed form by 
$$
p^{(t)}=\frac{1}{n}\sum_{i=1}^n\tau_{i1}^{(t)}.
$$
On the other hand, even in the complete-data case, the MLEs of $\mu_i$, $\sigma_i$ and $\nu_i$ ($i=1,2$) can only be found via numerical maximization. The two summands of (\ref{eq:Qfunc}) can be maximized separately, since they depend on different parameters, so that $\mu_j^{(t)},\sigma_j^{(t)},\gamma_j^{(t)},\nu_j^{(t)}$ are obtained as follows: 
\begin{equation}
(\mu_j^{(t)},\sigma_j^{(t)},\gamma_j^{(t)},\nu_j^{(t)})=\arg\max_{\boldsymbol{\theta}_i}\sum_{i=1}^n\tau_{ij}\log f_j(x_i;\boldsymbol{\theta}_j),\quad j=1,2.
\label{eq:Mnum}
\end{equation}
The function $\sum_{i=1}^n\tau_{ij}\log f_j(x_i;\boldsymbol{\theta}_j)$ is a weighted version of the skewed-$t$ log-likelihood function and can be maximized analogously, by means of standard optimization routines.

It is worth noting that estimating the number of degrees of freedom is a difficult task even in the basic $t$ distribution; see, e.g., \cite{lan89}, \cite{ven02} and \cite{vil14}. The latter paper describes the numerical issues related to the almost flat likelihood corresponding to large values of $\nu$: considering that the $t$ distribution converges to the normal as $\nu\to\infty$, \cite{vil14} propose to set a \emph{turning point} $\nu^*$, defined as the value of $\nu$ where a $t$ distribution ``turns'' into a normal \citep[p. 199]{vil14}. Since we have encountered some numerical difficulties in our simulations, especially regarding the asymptotic distribution of the test statistics (see Sect. \ref{sec:test}), we adopt the same approach here, setting $\nu^*=30$, analogously to \cite{vil14}. 

\subsection{Model selection and testing}
\label{sec:mod_sel}

The proposed model is very flexible, but this feature typically requires to pay a price, in terms of possible instability of parameter estimates. Hence, it is important to assess whether such a flexibility is necessary. Put it differently, one should check whether the population components are skewed and/or heavy-tailed. Heavy-tailedness is directly modeled by the parameter $\nu$, since a large value of $\nu$ implies an essentially normal distribution, and the turning point $\nu^*$ automatically takes it into account.

On the other hand, it is useful to test whether $\gamma_i$ is equal to one in the $i$-th population. The null hypothesis of no skewness in the $j$-th population is $H_0:\gamma_j=1$, to be tested versus $H_1:\gamma_j\neq 1$. To this aim, it is possible to exploit the MLE procedure used for estimating the parameters: in particular, since the maximized likelihood is a by-product of the EM-algorithm, we can use a likelihood ratio test. Notice that MLE of the restricted model corresponding to the null hypothesis is straightforward: it suffices to replace $\gamma_j^{(t)}$ by 1 in (\ref{eq:Qfunc}), (\ref{eq:Estep}) and in the $j$-th equation (\ref{eq:Mnum}), which is now maximized only with respect to $\mu_j^{(t)}$, $\sigma_j^{(t)}$ and $\nu_j^{(t)}$.

Without loss of generality, consider testing $H_0:\gamma_1=1$ and let $\ell(\boldsymbol{\theta};\boldsymbol{x})$ and $\ell_0(\boldsymbol{\theta}_0;\boldsymbol{x})$ be respectively the unrestricted and restricted log-likelihood functions, where $\boldsymbol{\theta}_0=(p,\mu_1,\sigma_1,\nu_1,\mu_2,\sigma_2,\gamma_2,\nu_2)'$. The log-likelihood ratio test is given by
\begin{equation}
\label{eq:llr}
\lambda=2(\ell(\hat{\boldsymbol{\theta}};\boldsymbol{x})-\ell_0(\hat{\boldsymbol{\theta}}_0;\boldsymbol{x})).
\end{equation}
Under regularity conditions, the asymptotic distribution of $\lambda$ is $\chi^2_k$, where $k$ is equal to the difference between the dimension of the unrestricted and restricted parameter space, so that, in the present case, $k=1$ \citep[p. 462-3]{leh98}. However, even though the regularity conditions are formally satisfied, the almost flatness of the $t$ log-likelihood, even for moderate $\nu$, is likely to be an issue \citep{vanzy15}. Hence, in the following we will double-check the null distribution via simulation.

\section{Simulation experiments}
\label{sec:sim}

\subsection{Well-specified setup}

The purpose of the experiments in this section is to study the finite-sample behavior of the estimators when the true data-generating process (DGP) is the two-population mixture of skewed-$t$ distributions with density (\ref{eq:mixdens}). 

The scenarios employed in the simulation are defined by the parameter values in Table \ref{tab:scenarios}; here and in the following, the three scenarios in each row correspond to the three values of $p$ reported in the second column. All experiments are performed with sample sizes $n\in\{100,500,1000\}$.

\begin{table}[htbp]
\centering
\caption{Well-specified case: the parameter values defining the 9 scenarios.}
\begin{tabular}{c|ccccccccc}
	\midrule
	Scenario	& $p$ & $\mu_1$ & $\sigma_1$ & $\gamma_1$ & $\nu_1$ & $\mu_2$ & $\sigma_2$ & $\gamma_2$ & $\nu_2$  \\
	\hline 
	1-3 & \{0.2,0.5,0.8\} & $-0.8$ & 1.3 & 0.8 & 4 & 0.8 & 1 & 1.3 & 15 \\
	\hline
	4-6	& \{0.2,0.5,0.8\} & $-1$ & 1.3 & 0.8 & 4 & 1 & 1 & 1.3 & 15 \\
	\hline
	7-9	& \{0.2,0.5,0.8\} & $-2$ & 1.3 & 0.8 & 4 & 2 & 1 & 1.3 & 15 \\
	\midrule
\end{tabular}
\label{tab:scenarios}
\end{table}

The choice of parameters in Table \ref{tab:scenarios} allows us to explore a wide variety of density shapes: scenarios 7-9 correspond to a bimodal distribution with well-separated populations, scenario 5 is also bimodal, with more overlap, finally the remaining parameter configurations yield a unimodal density. Notice also that, in each scenario, the first population is negatively skewed ($\gamma_1<1$) and heavy-tailed (``small'' $\nu_1$), whereas the second one has positive skewness  ($\gamma_2>1$) and light tails (``large'' $\nu_2$).

In addition to the skewed-$t$ mixture, we have also estimated a g-and-h mixture: in this case, the interest is only in the Kolmogorov-Smirnov (KS) test, to check whether the g-and-h mixture can yield a good fit in a mis-specified setup. The reverse analysis (i.e., the assessment of the fit of the skewed-$t$ mixture to data from a g-and-h mixture) will be carried out in the next subsection. All the outcomes displayed in the following  are based on $B=500$ replications. The null hypothesis is that two samples with the same sample size arise from a common unspecified distribution function, and the computation is carried out via the \texttt{ks.test} function. In addition to the KS test, we have also employed the two-sample Anderson-Darling test.

\begin{sidewaysfigure}
\centering
\includegraphics[angle=270,width=18cm]{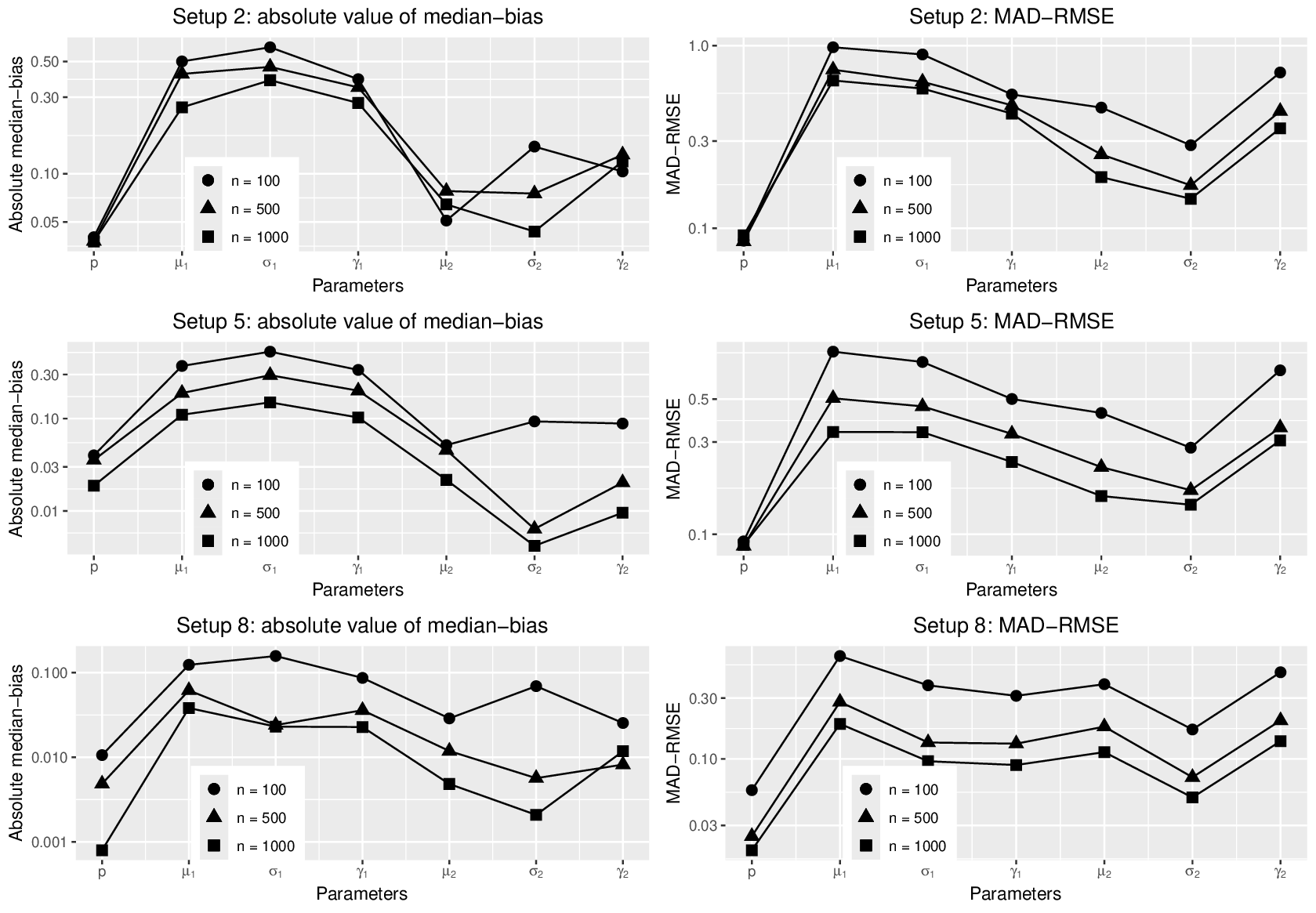}
\caption{Absolute median-bias and MAD-RMSE in setups 2, 5 and 8.}
\label{fig:b_RMSE}
\end{sidewaysfigure}

Figure \ref{fig:b_RMSE} shows the absolute value of the median-bias and the MAD-RMSE for three of the nine setups in Table \ref{tab:scenarios} (the results for the remaining scenarios are quite similar and therefore omitted). The median-bias and MAD-RMSE are defined as  $b_{med}(\hat{\theta}_j)\stackrel{\text{def}}{=}\text{median}(\hat{\theta}_j)-\theta_j$ and MAD-RMSE$=\sqrt{b_{med}(\hat{\theta})^2+\text{mad}(\hat{\theta})^2}$ respectively, where the mean absolute deviation (MAD) is given by $\text{mad}(\hat{\theta}_j)\stackrel{\text{def}}{=}(1/B)\sum_{i=1}^B|\hat{\theta}_{ij}-\text{median}(\hat{\theta}_j)|$. They are used in place of the more familiar (mean-based) bias and RMSE because, when $n=100$, in a few replications, possibly corresponding to cases where the algorithm did not converge, the estimates of $\gamma_i$ ($i=1,2$) are extremely large. On the other hand, when $n\in\{500,1000\}$, median-bias and MAD-RMSE are almost identical to usual bias and RMSE. Both plots are semilogarithmic, which is the reason why we display the absolute value of the bias. 

The main message arising from Figure \ref{fig:b_RMSE} is that there is a considerable improvement, both in terms of median-bias and in terms of MAD-RMSE, when the sample size is 500 instead of 100, whereas the advantage is smaller when $n$ is equal to 1000 instead of 500. Similarly, both the absolute bias and the RMSE decrease when moving from Setup 2 to 5 and from Setup 5 to 8, namely when the separation between the population increases, in line with well-known theoretical results about MLE of finite mixture distributions \citep{mcl00}.

Table \ref{tab:nuest} displays some numerical evidence about the estimates of $\nu_1$ and $\nu_2$: the first two columns report the medians, whereas the last two display the percentage of values smaller than 30, defined as perc$_{\nu_j}=:100\cdot \#\{\hat{\nu}_j<30\}/B$. The ``cap'' at 30 is effectively binding mostly when $n=100$, and especially for $\nu_2$, whose true value is larger; when $n>100$, not only $\nu_1$, but also $\nu_2$, is estimated rather well.

\begin{table}[htbp]
\centering
\caption{Estimates of $\nu_1$ and $\nu_2$, and percentage of replications where the estimates are smaller than 30; $\hat{\nu}_1$ and $\hat{\nu}_2$ are the median of all estimates.}
\begin{tabular}{c|ccccc}
\midrule
Scenario & $n$ & $\hat{\nu}_1$ & $\hat{\nu}_2$ & perc$_{\nu_1}$ & perc$_{\nu_2}$  \\
\midrule
\multirow{3}{*}{2} 	& 100 & 7.59 & 30.00 & 63.4   & 44.0  \\
& 500 & 5.02 & 15.42 & 93.1   & 60.8  \\
& 1000 & 4.52 & 13.32 & 98.6 & 67.2 \\
\hline
\multirow{3}{*}{5} 	& 100 & 6.94 & 30.00 & 65.8 & 42.2  \\
& 500 & 4.78 & 19.28 & 94.6 & 56.6  \\
& 1000 & 4.36 & 14.16 & 98.0 & 65.8 \\
\hline
\multirow{3}{*}{8} 	& 100 & 6.32 & 30.00 & 70.2  & 47.8  \\
& 500 & 4.40 & 15.79 & 97.4   & 60.8  \\
& 1000 & 4.22 & 15.44 & 99.8  & 69.8 \\
\midrule
\end{tabular}
\label{tab:nuest}
\end{table}

Finally, Table \ref{tab:pKS} displays the outcomes of the KS test for the estimated skewed-$t$ mixture and g-and-h mixtures with respect to the true skewed-$t$ mixture. More precisely, we use the two-sample KS test, i.e. a test for the null hypothesis that two samples come from the same unspecified distribution, and report the average $p$-value and the percentage of replications with $p$-value smaller than 5\%. Specifically, $p_{st}$ is the $p$-value of the test when the two samples are generated from the true and the estimated skewed-$t$ mixture; similarly, $p_{GH}$ is the $p$-value of the test when the two samples are generated from the true skewed-$t$ mixture and from the estimated g-and-h mixture. Only the results for scenarios 2, 5 and 8 are displayed in Table \ref{tab:pKS}, since the evidence arising from the remaining scenarios is almost identical.


\begin{table}[htbp]
\centering
\caption{KS test: average $p$-value ($\overline{p}$) and percentage of replications with $p$-value smaller than 5\% (perc$_{5\%}$)  when the true DGP is a skewed-$t$ mixture.}
\begin{tabular}{c|c|cc|cc|cc}
\midrule
Scenario & & \multicolumn{2}{c}{$n=100$} & \multicolumn{2}{c}{$n=500$} & \multicolumn{2}{c}{$n=1000$}  \\
\hline 
& & $p_{st}$ & $p_{GH}$ & $p_{st}$ & $p_{GH}$  & $p_{st}$ & $p_{GH}$   \\
\hline
\multirow{2}{*}{2} & $\overline{p}$ &0.971 & 0.661 &     0.978     & 0.680   &   0.984  &    0.653 \\
& perc$_{5\%}$ & 0 & 1.4 &  0 & 0.4 &   0  & 0.6 \\		
\hline 
\multirow{2}{*}{5} & $\overline{p}$ & 0.975  &    0.644   & 0.973  &  0.651  &  0.976  & 0.665 \\
& perc$_{5\%}$ & 0 &  1.4  & 0 & 1.2 &   0 &  0 \\		
\hline
\multirow{2}{*}{8} & $\overline{p}$ & 0.974    &  0.579   &   0.979    &  0.568    &  0.977  &    0.561  \\
& perc$_{5\%}$ & 0 & 5.0 &   0 & 6.0 &   0 &   4.2 \\		
\midrule
\end{tabular}
\label{tab:pKS}
\end{table}

Besides confirming that the estimated and true skewed-$t$ mixtures are not significantly different, the outcomes in Table \ref{tab:pKS} suggest that the g-and-h mixture also provides a good fit: even though the average $p$-value is smaller, the null hypothesis is hardly ever rejected.

\subsection{Mis-specified setup}

Next we simulate observations from a g-and-h mixture and estimate it by means of both the correctly specified g-and-h mixture and the mis-specified skewed-$t$ mixture; the main goal consists in assessing the ability of the latter to fit bimodal data from a different DGP. Following \cite{zha25}, a random variable $V$ distributed as a two-population g-and-h mixture can be defined via the cdf:
$$
P(V\le v)=pP(V^{(1)}_{a_1,b_1,g_1,h_1}\le v)+(1-p)P(V^{(2)}_{a_2,b_2,g_2,h_2}\le v),
$$
where $(a_1,b_i,g_i,h_i)'$ is the parameter vector of the $i$-th g-and-h population. Notice, however, that neither the pdf nor the cdf of the g-and-h distribution is known; see \cite{zha25} for details.

We summarize in Table \ref{tab:setupGH} the parameters used for the simulation. Also in this setup, we report only the outcomes for scenarios 2, 5 and 8, since no major difference is observed in the remaining cases.

\begin{table}[htbp]
\centering
\caption{Mis-specified case: the parameter values defining the 9 scenarios.}
\begin{tabular}{c|ccccccccc}
	\midrule
	Scenario	& $p$ & $a_1$ & $b_1$ & $g_1$ & $h_1$ & $a_2$ & $b_2$ & $g_2$ & $h_2$  \\
	\hline 
	1-3 & \{0.2,0.5,0.8\} & $-1$ & 1 & $-0.5$ & 0.3 & 1 & 1 & 0.7 & 0.2 \\
	\hline
	4-6	& \{0.2,0.5,0.8\} & $-1.5$ & 1 & $-0.5$ & 0.3 & 1.5 & 1 & 0.7 & 0.2 \\
	\hline
	7-9	& \{0.2,0.5,0.8\} & $-2$ & 1 & $-0.5$ & 0.3 & 2 & 1 & 0.7 & 0.2 \\
	\midrule
\end{tabular}
\label{tab:setupGH}
\end{table}

Table \ref{tab:pKSmis} shows the outcomes of the two-sample KS test; $p_{st}$ is the $p$-value of the test when the two samples are generated from the true g-and-h mixture and from the estimated skewed-$t$ mixture; similarly, $p_{GH}$ is the $p$-value of the test when the two samples are generated from the true and estimated g-and-h mixture.

\begin{table}[htbp]
\centering
\caption{KS test: average $p$-value ($\overline{p}$) and percentage of replications with $p$-value smaller than 5\% (perc$_{5\%}$) when the true DGP is a g-and-h mixture.}
\begin{tabular}{c|c|cc|cc|cc}
\midrule
Scenario      &                & \multicolumn{2}{c}{$n=100$} & \multicolumn{2}{c}{$n=500$} & \multicolumn{2}{c}{$n=1000$} \\ \hline
&                & $p_{st}$ &     $p_{GH}$     & $p_{st}$ &     $p_{GH}$     & $p_{st}$ &     $p_{GH}$      \\ \hline
\multirow{2}{*}{2} & $\overline{p}$ &  0.724   &      0.598       &  0.619   &      0.648       &  0.527   &       0.654       \\
&  perc$_{5\%}$  &   0.4    &       3.4        &   0.8    &       0.8       &   0.6    &        0.8        \\ \hline
\multirow{2}{*}{5} & $\overline{p}$ &  0.743   &      0.607       &  0.618   &      0.613       &  0.519   &       0.638       \\
&  perc$_{5\%}$  &   0.4    &       2.2        &   1.0    &       2.2       &   1.2    &        1.8        \\ \hline
\multirow{2}{*}{8} & $\overline{p}$ &  0.729   &      0.619       &  0.620   &      0.578       &  0.506   &       0.567       \\
&  perc$_{5\%}$  &   0.2    &       1.4        &   0.4    &       4.6       &   1.0    &        2.0        \\ \midrule
\end{tabular}
\label{tab:pKSmis}
\end{table}

There is no major difference between the two distributions, even though the skewed-$t$ mixture seems to perform slightly better for the two smallest sample sizes.
This result confirms the extreme flexibility of the skewed-$t$ mixture, considering that it is mis-specified in this experiment. 

\subsection{Computing times}
\label{sec:comp}

Average computing times (in seconds) for the two estimation methods are reported in Table \ref{tab:compTimeCombined}.
\begin{table}[htbp]
\centering
\caption{Average computing times (in seconds) under different true DGPs.}
\begin{tabular}{c|c|cc|cc|cc}
\midrule
True DGP & Scenario & \multicolumn{2}{c}{$n=100$} & \multicolumn{2}{c}{$n=500$} & \multicolumn{2}{c}{$n=1000$} \\
\hline
&  & St & GH & St & GH & St & GH \\
\hline
\multirow{3}{*}{Skewed-$t$}
& 7 & 21.45 & 3.05 & 21.17 & 3.00 & 22.33 & 3.32 \\
& 8 & 9.52  & 2.84 & 15.65 & 3.14 & 22.16 & 3.29 \\
& 9 & 11.36 & 2.85 & 18.66 & 3.15 & 33.91 & 3.53 \\
\hline
\multirow{3}{*}{g-and-h}
& 7 & 7.19 & 3.37 & 7.17 & 3.38 & 9.68 & 3.61 \\
& 8 & 2.05 & 3.03 & 2.49 & 8.16 & 4.27 & 3.68 \\
& 9 & 7.17 & 3.02 & 8.16 & 3.40 & 17.48 & 3.78 \\
\midrule
\end{tabular}
\label{tab:compTimeCombined}
\end{table}
The outcomes show that estimation of the skewed-$t$ mixture takes longer, a result which is not surprising, since the EM algorithm converges linearly, with rate of convergence related to the amount of ``missing information'' \citep[Sect. 3.9.3]{mcl08}. The estimation of the skewed-$t$ mixture is faster in the mis-specified than in the correctly-specified setup. Upon double-checking the estimation results, we found that the algorithm converges in a much smaller number of iterations when the DGP is mis-specified. This outcome is likely to depend on the different shape of the log-likelihood function in the two setups, hence it is difficult to draw any general conclusion.

\subsection{Testing for skewness}
\label{sec:test}

\subsubsection{Null distribution of the test}

\begin{figure} 
\centering
\includegraphics[width=13cm]{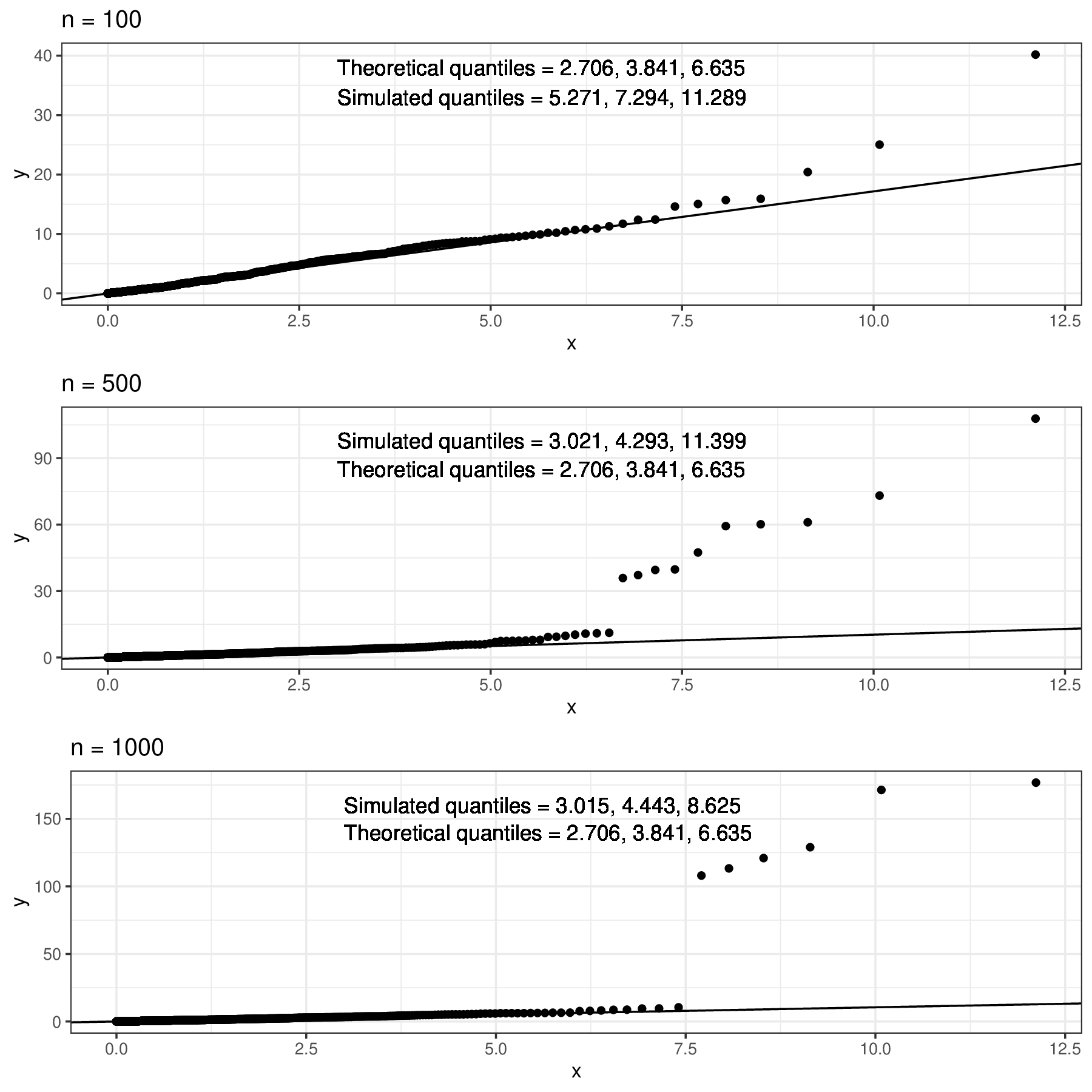} 
\caption{QQ-plots of the simulated null distribution of the test for $H_0:\gamma_1=1$ when $n\in\{100,500,1000\}$ vs. the theoretical $\chi^2_1$ distribution.}
\label{fig:null}
\end{figure}

In light of the remarks about the asymptotic distribution of the test statistic (\ref{eq:llr}) put forward in Sect. \ref{sec:mod_sel}, we simulate the null distribution of  $\lambda$. 
The QQ-plot of the simulated null distribution of the test vs. the theoretical $\chi^2_1$ distribution, namely the asymptotic null distribution of the test suggested by the theory, provided the regularity conditions are satisfied, is displayed in Figure \ref{fig:null}; the panels correspond to $n\in\{100,500,1000\}$ and also report the numerical values of the simulated quantiles at levels 90, 95 and 99\%. As expected, the $\chi^2_1$ approximation works better for larger sample sizes, however the tail behavior seems to remain slightly different. Given also the possible issues with the $t$ log-likelihood mentioned in Sect. \ref{sec:mod_sel}, in the following we will compute critical values from the simulated null distribution.

\subsubsection{Power of the test}

In this section we study the power function via simulation. At each replication of the experiment, we simulate a sample from (\ref{eq:mixdens}) with parameters $p = .5$, $\mu_1=-2$, $\sigma_1 = 1.2$, $\nu_1 = 4$, $\mu_2 = 2$, $\sigma_2 = 0.8$, $\gamma_2 = 1.3$, $\nu_2 = 15$, $\gamma_1\in\{1.1,1.2,\dots,2\}$, compute the test and make the decision at the $\alpha$ level, where $\alpha\in\{0.1,0.05,0.01\}$. After repeating this procedure $B=500$ times, we evaluate the power as
$$
\text{pow}_{\alpha}=\frac{\#\lambda:\lambda>cr_{\alpha}}{B},
$$
where $cr_{\alpha}$ is the simulated critical value at level $\alpha$ reported in Figure \ref{fig:null}. Table \ref{tab:power_combined} reports the outcomes.
\begin{table}[htbp]
\scriptsize 
\centering
\caption{Power of the test for different sample sizes $n$.}
	\begin{tabular}{c|c|cccccccccc}
		\midrule
		$n$ &  & \multicolumn{10}{c}{$\gamma_1$} \\
		\cmidrule(lr){3-12}
		&  & 1.1 & 1.2 & 1.3 & 1.4 & 1.5 & 1.6 & 1.7 & 1.8 & 1.9 & 2 \\
		\midrule
		\multirow{3}{*}{100}
		& $\alpha=0.9$  & 0.081 & 0.065 & 0.077 & 0.106 & 0.110 & 0.113 & 0.171 & 0.176 & 0.160 & 0.180 \\
		& $\alpha=0.95$ & 0.035 & 0.036 & 0.036 & 0.044 & 0.054 & 0.062 & 0.101 & 0.113 & 0.117 & 0.119 \\
		& $\alpha=0.99$ & 0.014 & 0.016 & 0.022 & 0.025 & 0.024 & 0.028 & 0.038 & 0.039 & 0.030 & 0.036 \\
		\midrule
		\multirow{3}{*}{500}
		& $\alpha=0.9$  & 0.120 & 0.172 & 0.299 & 0.312 & 0.447 & 0.406 & 0.576 & 0.621 & 0.664 & 0.694 \\
		& $\alpha=0.95$ & 0.056 & 0.096 & 0.185 & 0.219 & 0.347 & 0.256 & 0.465 & 0.479 & 0.570 & 0.579 \\
		& $\alpha=0.99$ & 0.009 & 0.014 & 0.014 & 0.030 & 0.055 & 0.056 & 0.055 & 0.050 & 0.119 & 0.170 \\
		\midrule
		\multirow{3}{*}{1000}
		& $\alpha=0.9$  & 0.144 & 0.221 & 0.479 & 0.585 & 0.626 & 0.731 & 0.754 & 0.850 & 0.880 & 0.875 \\
		& $\alpha=0.95$ & 0.081 & 0.149 & 0.295 & 0.461 & 0.494 & 0.618 & 0.647 & 0.763 & 0.814 & 0.862 \\
		& $\alpha=0.99$ & 0.011 & 0.035 & 0.057 & 0.178 & 0.200 & 0.287 & 0.342 & 0.426 & 0.475 & 0.644 \\
		\midrule
	\end{tabular}
\label{tab:power_combined}
\end{table}

In line with the theory, for a given sample size the power of the test increases when $\gamma_1$ gets larger; for the same value of $\gamma_1$, the power increases with the sample size. However, the sample size plays a major role: when $n=100$ the power remains quite low for all values of $\gamma_1$, whereas for $n=500$ and especially for $n=1000$ it increases sharply. 

\section{Empirical analysis}
\label{sec:emp}

\subsection{Shiller's Standard \& Poor's 500 distortion}
\label{sec:dist}

The distortion of the S\&P 500 index is defined as the log difference between its real stock market index and its real fundamental value \citep{schmi17}. The concept has been popularized by Nobel prize winner Robert Shiller \citep{shi15}, who also details how to compute the S\&P 500's fundamental value and makes available the updated data\footnote{\texttt{http://www.econ.yale.edu/~shiller/data.htm}}. The dataset starts in January 1871, and we use all data until August 2025, which means 1857 monthly observations of the real S\&P 500 and its dividend payments. The real fundamental value of the S\&P 500 is obtained by discounting its dividend payments, assuming a constant real discount rate and growth rate of the last observed dividend. Previous empirical analysis suggests that the distribution of the S\&P 500 distortion is bimodal, which would imply that the S\&P 500 spends relatively more time in bull and bear markets than near its fundamental value.

\begin{figure} 
\centering
\includegraphics[width=13cm]{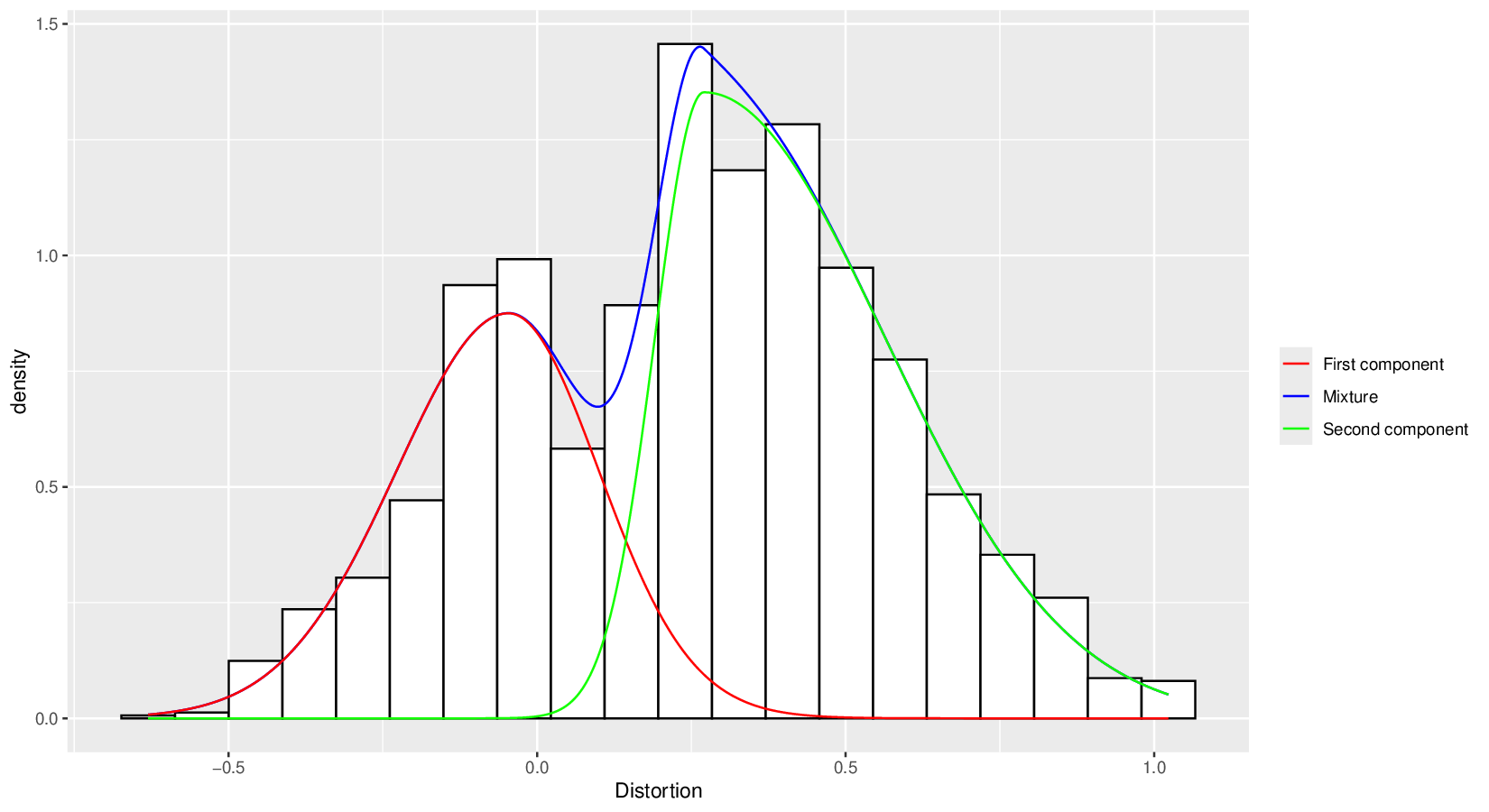}
\caption{The S\&P 500 distortion with the estimated skewed-$t$ density and the two component skewed-$t$ densities.}
\label{fig:distHist}
\end{figure}

\begin{figure} 
\centering
\includegraphics[width=13cm]{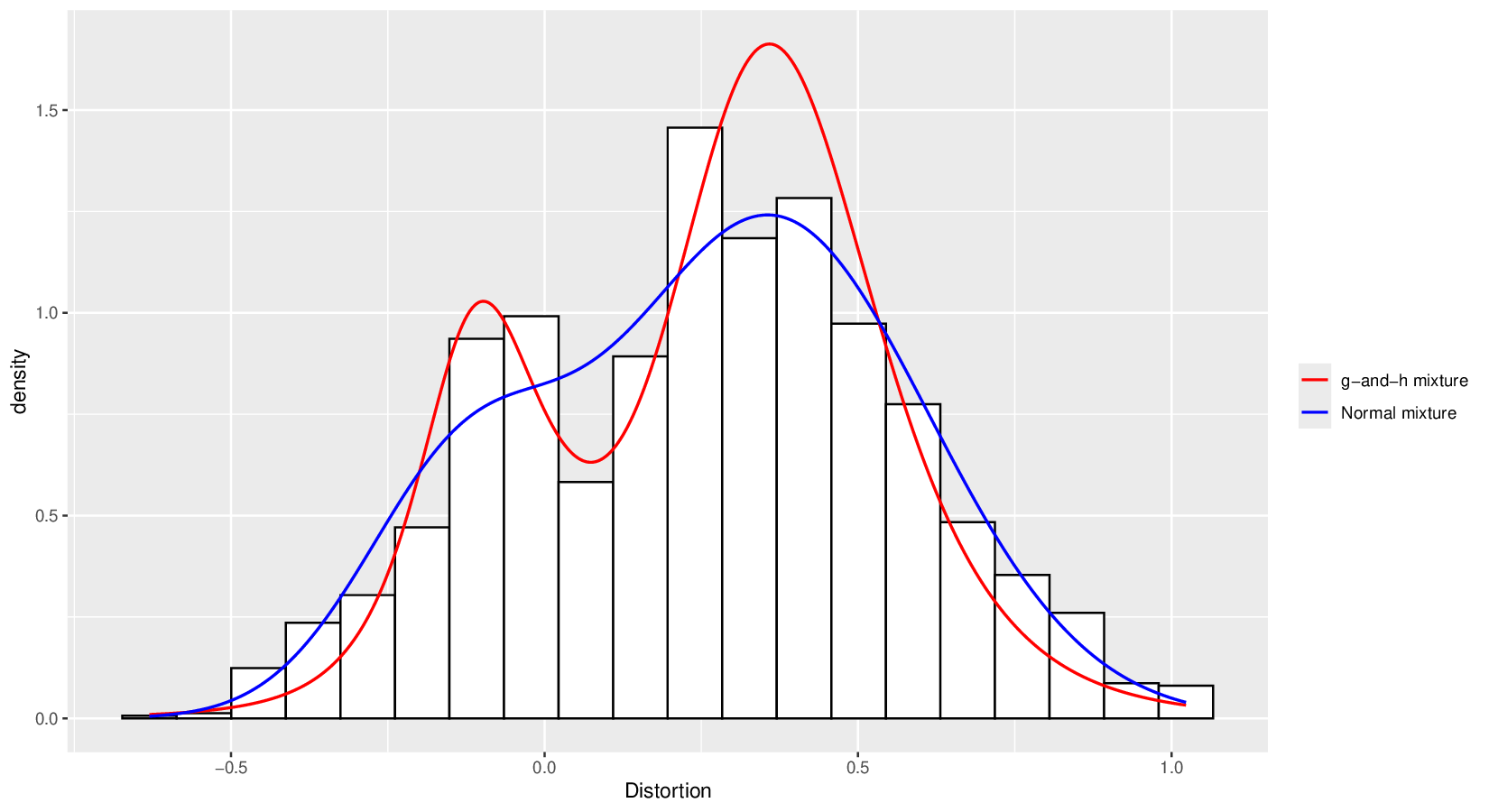}
\caption{The S\&P 500 distortion with the estimated g-and-h mixture and normal mixture.}
\label{fig:distHist1}
\end{figure}

The histogram of the distortion is shown in Figure \ref{fig:distHist}, along with the estimated skewed-$t$ mixture and the two skewed-$t$ component distributions. 
As can be seen from Figure \ref{fig:distHist}, the estimated density is bimodal. Figure \ref{fig:distHist1} shows that, when we fit g-and-h mixture or a normal mixture, the evidence may be different: the g-and-h mixture is bimodal, whereas the normal mixture is not. 
Moreover, the g-and-h mixture density seems to overestimate the main mode.

\FloatBarrier

To perform a thorough statistical investigation of the problem, we further explore the goodness of fit of the three approaches. The analysis is based on non-parametric bootstrap: for each model, at the $b$-th bootstrap replication we implement the following steps:
\begin{itemize}
\item checking whether the resulting mixture is bimodal;
\item computing the KS and AD tests.
\end{itemize}
These two steps are repeated $B=1000$ times: at each replication, a new bootstrap sample is employed to estimate the parameters and to compute the KS test; furthermore, the estimated density is checked for bimodality. At the end of the procedure we estimate the standard errors by means of the empirical standard deviation of the $B$ estimates.

For univariate normal mixtures, bimodality can be checked via the conditions of \cite{rob69}, which determine when the mixture is bimodal or unimodal on the basis of the numerical values of the parameters: see \citet[Example 5.5.1]{tit85}. In the two remaining cases, we assess bimodality numerically as follows: 
\begin{enumerate}
\item let $x_1,\dots,x_n$ be the observed sample, and $x_{(1)}\le \cdots\le x_{(n)}$ the corresponding order statistics;
\item evaluate the estimated density $\hat{f}$ over a grid of points $w_1<\cdots<w_m$ such that $w_1=x_{(1)}$ and $w_m=x_{(n)}$;
\item compute the first differences $\widehat{df}_i=\hat{f}(w_i)-\hat{f}(w_{i-1})$, $i=2,\dots,m$;
\item compute the signs of the sequence $\widehat{df}_i$, $i=2,\dots,m$ and the corresponding runs $ru$;
\item if $ru=2$, with signs ``-'' and ``+'', the density is unimodal;
\item if $ru=4$, with signs ``-'', ``+'', ``-'' and ``+'', the density is bimodal.
\end{enumerate}
Provided the grid is not too coarse, this procedure gives a rather accurate result, possibly missing only tiny bimodalities. Note that this approach is similar to the forward-difference method used for numerical approximation of derivatives \citep[Sect. 4.1]{bur10}.

Tables \ref{tab:app1_t}, \ref{tab:app1_gh} and  \ref{tab:app1_norm} show parameter estimates, standard errors, mean absolute deviations, average $p$-values ($\overline{p}$) of the KS test, percentage of $p$-values smaller than 5\% and percentage of cases where the estimated density is bimodal (perc$_{bim}$). 

\begin{table}[htbp]
\tiny
\centering
\caption{Parameter estimates (PE), standard errors (SE), mean absolute deviation (MAD), $p$-value of the KS and AD tests ($p_{KS}^{obs}$ and $p_{AD}^{obs}$), average $p$-values ($\overline{p}$) of the KS test over the $B$ bootstrap replications, percentage of KS $p$-values smaller than 5\% (perc$_{5\%}$) and percentage of bimodal densities for the skewed-$t$ mixture.}
	\setlength{\tabcolsep}{3pt}
	\begin{tabular}{c|cccccccccccccc}
		\midrule
		& $p$ & $\mu_1$ & $\sigma_1$ & $\gamma_1$ & $\nu_1$ & $\mu_2$ & $\sigma_2$ & $\gamma_2$ & $\nu_2$ & $p_{KS}^{obs}$ & $p_{AD}^{obs}$ & $\overline{p}$ & perc$_{5\%}$ & perc$_{bim}$ \\
		\hline
		PE & 0.366 &  $-0.046$   & 0.165 & 0.899 & 51.17 &  0.270  & 0.153 & 1.922 &  $+\infty$ & \multirow{3}{*}{0.638} & \multirow{3}{*}{0.249} & \multirow{3}{*}{0.342} & \multirow{3}{*}{11.7} & \multirow{3}{*}{99.3} \\ 
		SE & 0.029 & 0.016 & 0.021 & 0.098 & - & 0.021 & 0.012 & 0.202 & - & & & & & \\
		MAD & 0.026 & 0.012 & 0.017 & 0.082 & - & 0.017 & 0.009 & 0.137 & - & & & & & \\ \midrule
	\end{tabular}
\label{tab:app1_t}
\end{table}

\begin{table}[htbp]
\tiny
\centering
\caption{Parameter estimates (PE), standard errors (SE), mean absolute deviation (MAD), $p$-value of the KS and AD tests ($p_{KS}^{obs}$ and $p_{AD}^{obs}$), average $p$-values ($\overline{p}$) of the KS test, percentage of KS $p$-values smaller than 5\% (perc$_{5\%}$) and percentage of bimodal densities for the g-and-h mixture.}
	\setlength{\tabcolsep}{3pt}
	\begin{tabular}{c|cccccccccccccc}
		\midrule
		& $p$ & $a_1$ & $b_1$ & $g_1$ & $h_1$ & $a_2$ & $b_2$ & $g_2$ & $h_2$ & $p_{KS}^{obs}$ & $p_{AD}^{obs}$ & $\overline{p}$ & perc$_{5\%}$ &  perc$_{bim}$ \\
		\hline
		PE & 0.327 &  $-0.078$   & 0.134 & 0.309 & 0.253 &  0.385  & 0.169 & 0.156 &  0.066 & \multirow{3}{*}{0.082} & \multirow{3}{*}{0.012} & \multirow{3}{*}{0.228} & \multirow{3}{*}{26.6} & \multirow{3}{*}{58.7} \\ 
		SE & 0.064 & 0.030 & 0.035 & 0.348 & 0.133 & 0.027 & 0.023 & 0.069 & 0.020 & & & & &  \\ 
		MAD & 0.055 & 0.022 & 0.032 & 0.467 & 0.018 & 0.024 & 0.027 & 0.068 & 0.006 & & & & &  \\ \midrule
	\end{tabular}
\label{tab:app1_gh}
\end{table}

Tables \ref{tab:app1_t} and \ref{tab:app1_gh} suggest that the fit of the skewed-$t$ mixture is overall better: $\overline{p}$ is larger, and perc$_{5\%}$ is significantly smaller. It should, however, be noticed that the parameters $g_1$, $g_2$ and $h_2$ in Table \ref{tab:app1_gh} are non-significant, so that the use of a mixture with g-and-h components would not be justified. The skewed-$t$ mixture is almost always bimodal, whereas for the g-and-h the percentage of bimodal estimated densities is only about 59\%. Since the distribution that seems to fit the data best is also the one with the largest perc$_{bim}$, bimodality of the distribution is confirmed.

\begin{table}[htbp]
\centering
\scriptsize
\caption{Parameter estimates (PE), standard errors (SE), mean absolute deviation (MAD), $p$-value of the KS and AD tests ($p_{KS}^{obs}$ and $p_{AD}^{obs}$), average $p$-values ($\overline{p}$) of the KS test, percentage of KS $p$-values smaller than 5\% (perc$_{5\%}$) and percentage of bimodal densities for the normal mixture.}
	\begin{tabular}{c|cccccccccc}
		\midrule
		& $p$ & $\mu_1$ & $\sigma_1$ & $\mu_2$ & $\sigma_2$ & $p_{KS}^{obs}$ & $p_{AD}^{obs}$ & $\overline{p}$ & perc$_{5\%}$ & perc$_{bim}$ \\
		\hline
		PE & 0.218 &  $-0.135$   & 0.160 & 0.361  &  0.252 & \multirow{3}{*}{0.646} & \multirow{3}{*}{0.414} & \multirow{3}{*}{0.227} & \multirow{3}{*}{18.4} & \multirow{3}{*}{10.3}\\ 
		SE & 0.026 & 0.015 & 0.010 & 0.013 & 0.009 & & & & & \\ 
		MAD & 0.026 & 0.014 & 0.009 & 0.013 & 0.009 & & & & & \\ \midrule
	\end{tabular}
\label{tab:app1_norm}
\end{table}

Finally, Table \ref{tab:app1_norm} displays the results of the same analysis carried out with normal mixtures. In terms of goodness of fit, the normal and skewed-$t$ mixtures are comparable, since the average KS $p$-value is approximately the same, and the test rejects the null hypothesis that two samples arise from a common unspecified distribution function with a similar frequency. As for bimodality, we notice that the estimated normal mixture is mostly unimodal.

\begin{figure} 
\centering
\includegraphics[width=13cm]{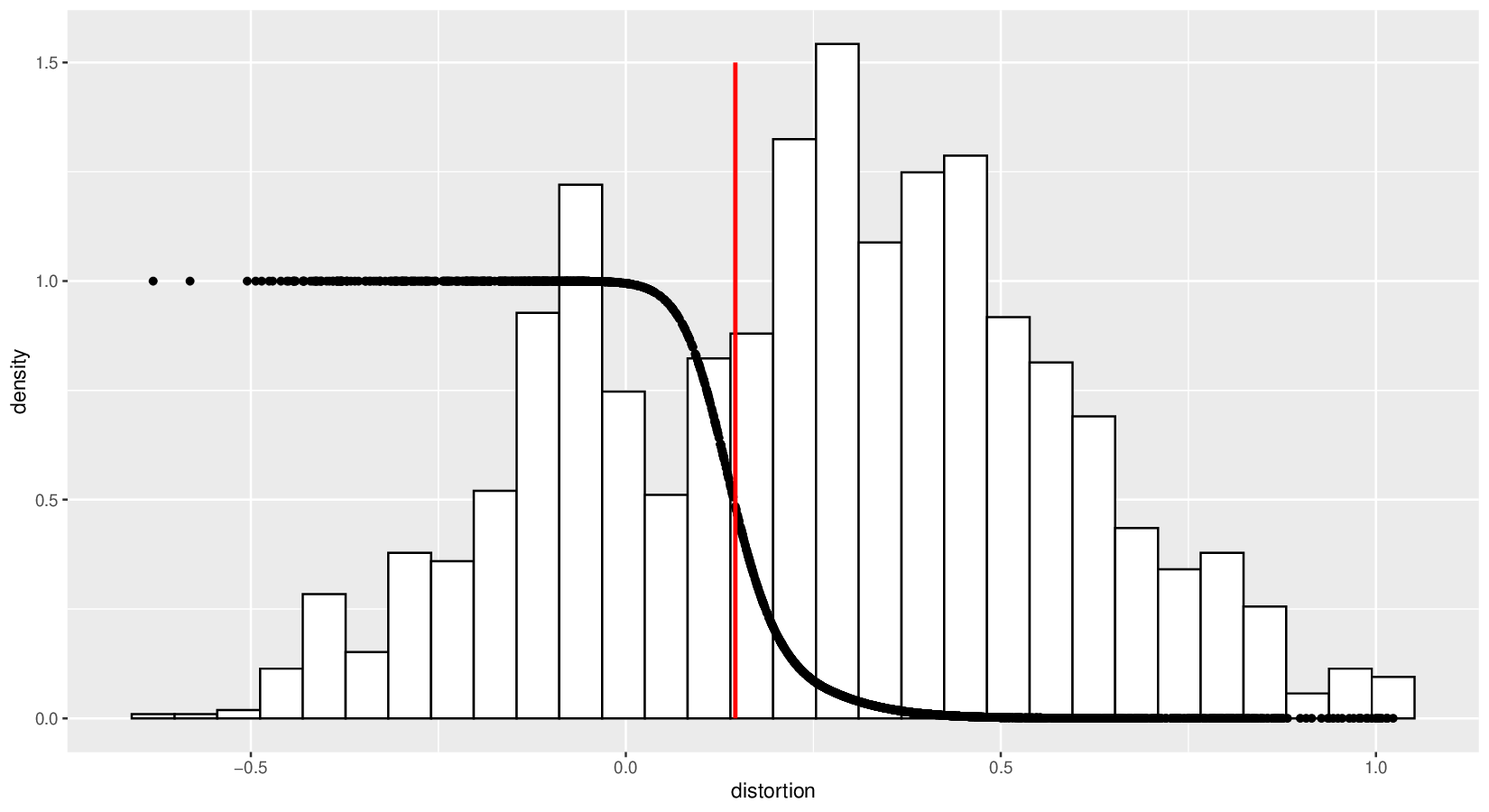}
\caption{Shiller's distortion (histogram), estimated posterior probabilities of the first population (scatterplot) and classification cutoff (vertical line).}
\label{fig:postdist1}
\end{figure}

Classification of bear and bull markets can be based on posterior probabilities. As usual, the classification cutoff is based on the estimated posterior probabilities: if $\tau_{i1}$ is larger (smaller) than 0.5, the $i$-th observation belongs to the first (second) population. Figure \ref{fig:postdist1} shows the histogram and the posterior probabilities of belonging to the first population (bear market).
\FloatBarrier

The cutoff, represented by the red vertical line, corresponds to the 650-th ordered observation, equal to 0.146; in other words, the 650 observations smaller than 0.146 are classified in the first population, corresponding to bear market conditions, the remaining observations in the second (bull market conditions). 
Overall, the empirical evidence displayed in this section allows us to conclude, in line with \cite{schmi17}, that the S\&P500 ``spends relatively more time in bull and bear markets than in the vicinity of its fundamental value''.

\subsection{Stamps thickness}

\begin{figure} 
\centering
\includegraphics[width=13cm]{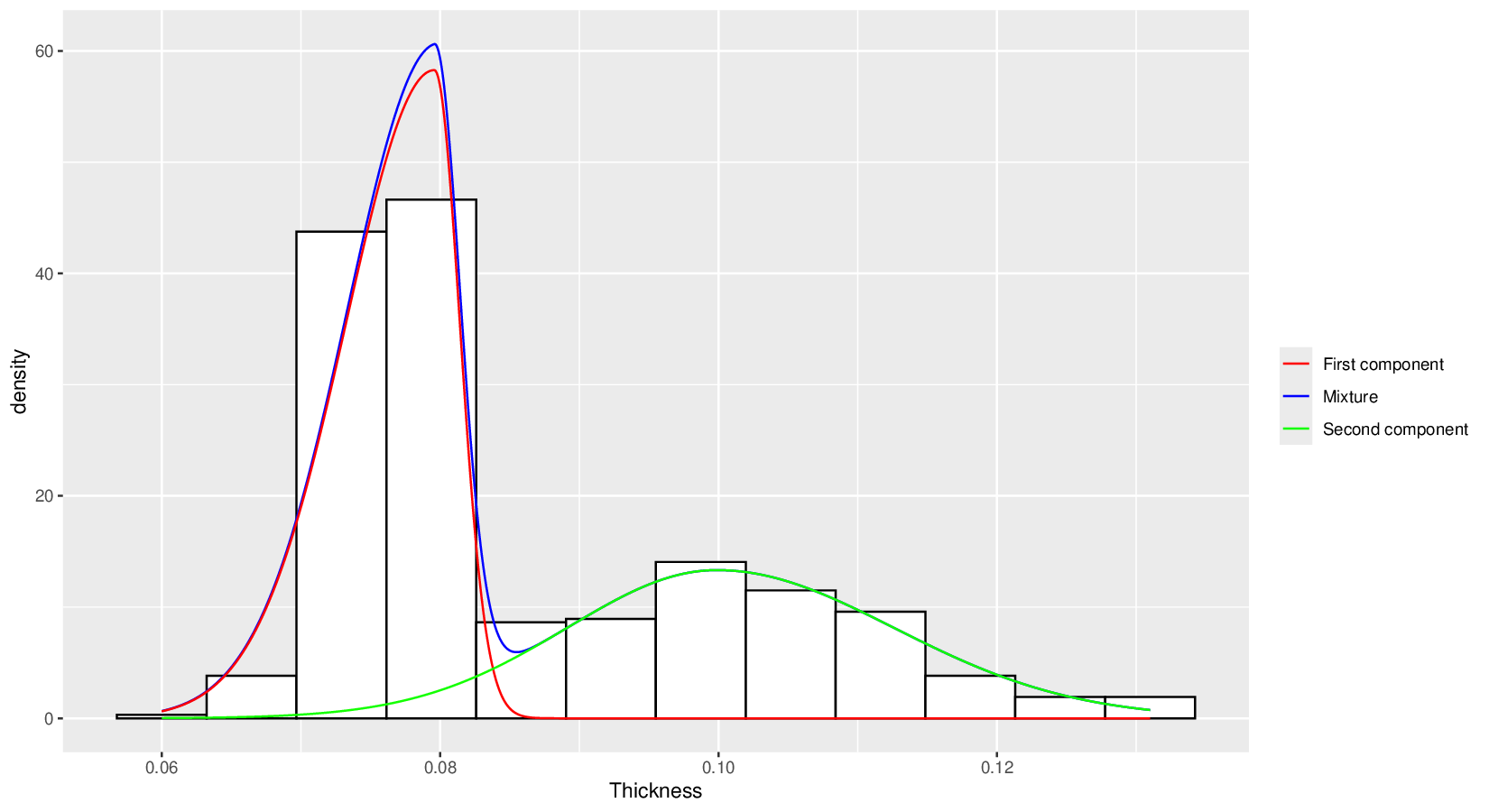}
\caption{The stamps thickness data with the estimated skewed-$t$ mixture density and the two component densities.}
\label{fig:distHistb}
\end{figure}

\begin{figure} 
\centering
\includegraphics[width=13cm]{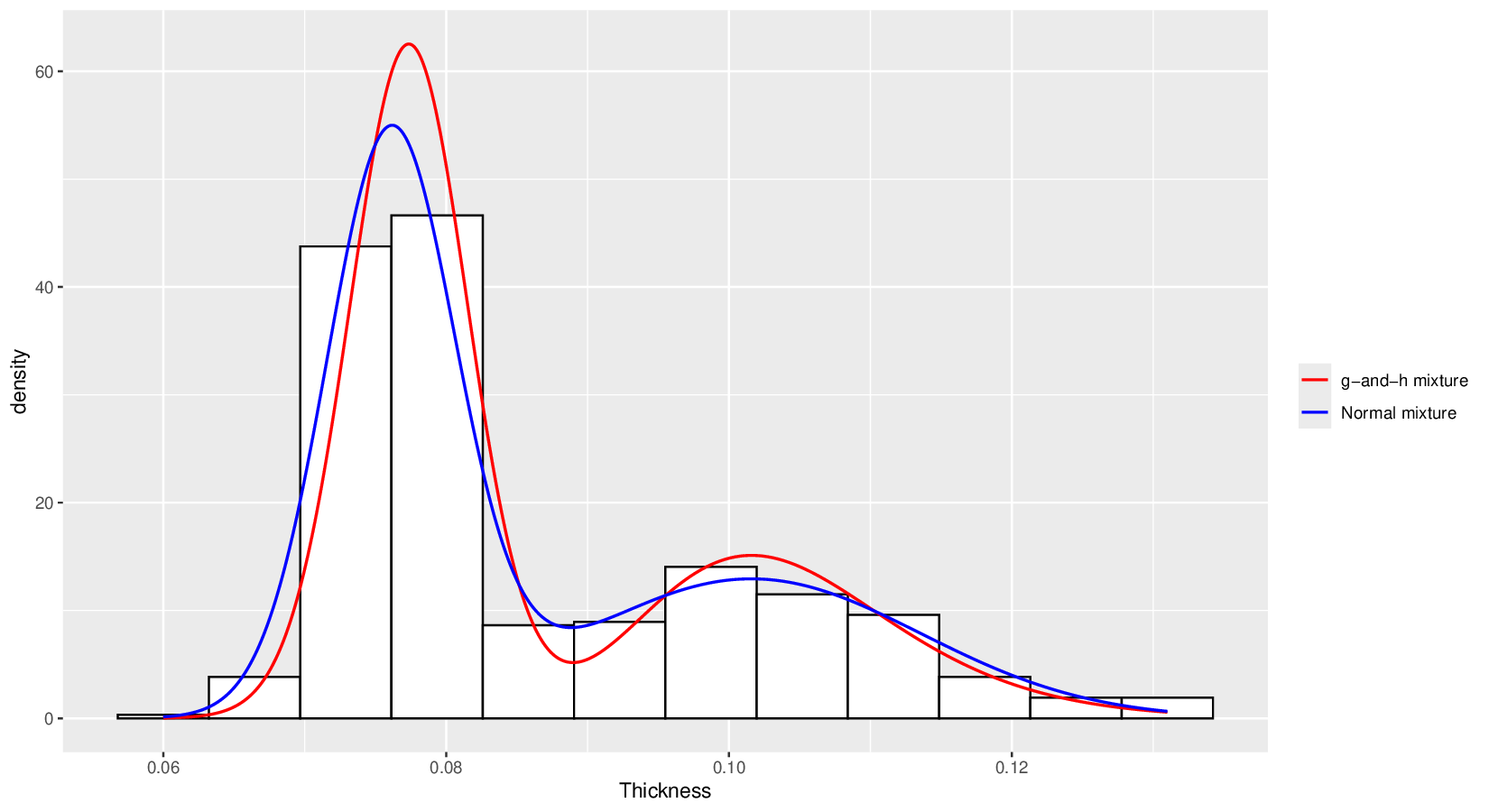}
\caption{The stamp thickness data with the estimated skewed-$t$ density and the two component skewed-$t$ densities.}
\label{fig:distHistb1}
\end{figure}

The dataset employed in this section is available in the \texttt{multimode} \texttt{R} package. It contains thickness measurements (in millimeters) of 485 unwatermarked used white wove stamps of the 1872 Hidalgo stamp issue of Mexico, and was first analyzed in \cite{ize88} and \cite{ame19}.
Figure \ref{fig:distHistb} shows the data as well as the fitted skewed-$t$ density. Similarly, Figure \ref{fig:distHistb1} displays the g-and-h and normal mixture densities. Tables \ref{tab:app2_t}, \ref{tab:app2_gh} and \ref{tab:app2_norm} show parameter estimates, standard errors and outcomes of the goodness-of-fit tests.

\begin{table}[htbp]
\tiny
\centering
\caption{Parameter estimates (PE), standard errors (SE), mean absolute deviation (MAD), $p$-value of the KS and AD tests ($p_{KS}^{obs}$ and $p_{AD}^{obs}$), average $p$-values ($\overline{p}$) of the KS test over the $B$ bootstrap replications, percentage of KS $p$-values smaller than 5\% (perc$_{5\%}$) and percentage of bimodal densities for the skewed-$t$ mixture.}
	\setlength{\tabcolsep}{3pt}
	\begin{tabular}{c|cccccccccccccc}
		\midrule
		& $p$ & $\mu_1$ & $\sigma_1$ & $\gamma_1$ & $\nu_1$ & $\mu_2$ & $\sigma_2$ & $\gamma_2$ & $\nu_2$ & $p_{KS}^{obs}$ & $p_{AD}^{obs}$ & $\overline{p}$ & perc$_{5\%}$ & perc$_{bim}$ \\
		\hline
		PE & 0.603 &  0.080  & 0.003 & 0.537 & 30 & 0.100 &  0.012  & 1.088 & 30 &  \multirow{3}{*}{0.160} & \multirow{3}{*}{0.221} & \multirow{3}{*}{0.067} & \multirow{3}{*}{54.2} & \multirow{3}{*}{98.9} \\ 
		SE & 0.035 & 0.002 & 0.001 & 0.886 & - & 0.005 & 0.003 & 19.380 & - & & & & & \\ 
		MAD & 0.029 & $<0.001$ & $<0.001$ & 0.060 & - & 0.002 & 0.002 & 0.143 & - & & & & & \\ \midrule
	\end{tabular}
\label{tab:app2_t}
\end{table}

\begin{table}[htbp]
\tiny
\centering
\caption{Parameter estimates (PE), standard errors (SE), mean absolute deviation (MAD), $p$-value of the KS and AD tests ($p_{KS}^{obs}$ and $p_{AD}^{obs}$), average $p$-values ($\overline{p}$) of the KS test, percentage of KS $p$-values smaller than 5\% (perc$_{5\%}$) and percentage of bimodal densities for the g-and-h mixture.}
	\setlength{\tabcolsep}{3pt}
	\begin{tabular}{c|cccccccccccccc}
		\midrule
		& $p$ & $a_1$ & $b_1$ & $g_1$ & $h_1$ & $a_2$ & $b_2$ & $g_2$ & $h_2$ & $p_{KS}^{obs}$ & $p_{AD}^{obs}$ & $\overline{p}$ & perc$_{5\%}$ &  perc$_{bim}$ \\
		\hline
		PE & 0.671 &  0.077   & 0.004 & 0.024 & 0.007 &  0.104  & 0.009 & 0.240 &  0.008 & \multirow{3}{*}{0.002} & \multirow{3}{*}{0.002} & \multirow{3}{*}{0.006} & \multirow{3}{*}{98.3} & \multirow{3}{*}{54.3} \\ 
		SE & 0.036 & 0.001 & 0.001 & 0.139 & 0.002 & 0.002 & 0.002 & 0.141 & 0.017 & & & & &  \\
		MAD & 0.020 & 0.001 & $<0.001$ & 0.084 & 0.001 & 0.002 & 0.001 & 0.098 & 0.001 & & & & &  \\ \midrule
	\end{tabular}
\label{tab:app2_gh}
\end{table}

\begin{table}[htbp]
\tiny
\centering
\caption{Parameter estimates (PE), standard errors (SE), mean absolute deviation (MAD), $p$-value of the KS and AD tests ($p_{KS}^{obs}$ and $p_{AD}^{obs}$), average $p$-values ($\overline{p}$) of the KS test, percentage of KS $p$-values smaller than 5\% (perc$_{5\%}$) and percentage of bimodal densities for the normal mixture.}
	\begin{tabular}{c|cccccccccc}
		\midrule
		& $p$ & $\mu_1$ & $\sigma_1$ & $\mu_2$ & $\sigma_2$ & $p_{KS}^{obs}$ & $p_{AD}^{obs}$ & $\overline{p}$ & perc$_{5\%}$ & perc$_{bim}$ \\
		\hline
		PE & 0.610 &  0.076   & 0.005 & 0.102  &  0.012 & \multirow{3}{*}{0.119} & \multirow{3}{*}{0.219} & \multirow{3}{*}{0.042} & \multirow{3}{*}{69.5} & \multirow{3}{*}{100}\\ 
		SE & 0.030 & $<0.001$  & 0.001 & $<0.001$ & 0.001 & & & & & \\
		MAD & 0.030 & $<0.001$  & 0.001 & $<0.001$ & 0.001 & & & & & \\ \midrule
	\end{tabular}
\label{tab:app2_norm}
\end{table}
According to the tables, the skewed-$t$ mixture yields the best fit, in terms of the observed values of the KS and AD tests, of the average KS $p$-value and of the percentage of KS $p$-values smaller than 5\%. The large standard errors of the $\hat{\gamma}_i$s are presumably due to the a few cases where the algorithm did not converge, as witnessed by the much smaller mean absolute deviation.

On the other hand, the g-and-h mixture is the worst model, but, similarly to the previous application, the parameters $g_1$, $g_2$, $h_1$, $h_2$ in Table \ref{tab:app2_gh} are all non-significant. In principle, this suggests that a normal mixture might be good enough, so that it is not surprising that the normal mixture fit is reasonably good. However, it is not as good as the skewed-$t$ mixture. As a possible explanation, in the latter model the first population is negatively skewed ($\hat{\gamma}_1$ well below 1), a feature that the normal mixture cannot capture. We conjecture that the ability of taking into account this negative skewness improves the fit of the skewed-$t$ mixture.

Finally, the two models characterized by the best goodness-of-fit measures, namely the skewed-$t$ and the normal mixture, are almost always bimodal. Hence, bimodality seems to be confirmed.

\section{Conclusion}
\label{sec:concl}

We investigate the use of a two-population mixture of skewed-$t$ distributions as a flexible tool for modeling bimodal, skewed and heavy-tailed datasets. We develop an EM algorithm for parameter estimation and a likelihood ratio test for the null hypothesis of no skewness in one of the components. Overall, the evidence provided by both simulation experiments and real-data analyses suggests that the proposed model is very flexible. With respect to a recently proposed mixture of g-and-h distributions, the fit seems better, even in mis-specified setups.

The model can be easily extended by using component distributions based on skewed versions of other symmetric distributions, which can be easily obtained via (\ref{eq:skd}). Further research would be necessary in order to investigate the properties and the estimation of these distributions.

\bigskip
\noindent\textbf{Acknowledgements} We thank prof. F. Westerhoff for providing the code for the computation of distortion in Section \ref{sec:dist}.

\section*{Declarations}

\begin{itemize}
\item Funding: No funding was received.
\item The authors report there are no competing interests to declare.
\item Data availability: the S\&P 500 distortion from January 1871 until August 2025 can be found in the \texttt{R} package \texttt{stMix}, available at\\ \texttt{https://github.com/marco-bee/stMix}. The stamps thickness dataset can be downloaded from the \texttt{multimode} \texttt{R} package.
\item Code availability: the \texttt{R} package \texttt{stMix}, containing the codes that implement the methods developed in this paper, is available at\\ \texttt{https://github.com/marco-bee/stMix}.
\item The authors contributed equally to this work.	
\item The authors report generative AI was not used in their research or preparation of this manuscript.
\end{itemize}

\begin{appendix}
\renewcommand{\theequation}{A-\arabic{equation}}
\setcounter{equation}{0}  
\setcounter{table}{0}
\renewcommand{\thetable}{A\arabic{table}}
\section{Proofs}	
\label{sec:app}

\subsection{Proof of Theorem \ref{teo:teo1}}

Let first $x<0$. From (\ref{eq:skd}) we have:
\begin{align}
	\label{eq:cdf1}
	F_{ST}(x;\gamma,\nu)&=\int_{-\infty}^xf_{ST}(t;\gamma,\nu)dt=\frac{2}{\gamma+\frac{1}{\gamma}}\int_{-\infty}^x \tilde{f}_{ST}(\gamma t;\nu)dt=\nonumber \\
	&=\frac{2}{\gamma+\frac{1}{\gamma}}\int_{-\infty}^{x\gamma} \tilde{f}_{ST}(u;\nu)\frac{1}{\gamma}du=\frac{2}{\gamma^2+1}\int_{-\infty}^{x\gamma} \tilde{f}_{ST}(u;\nu)du=\nonumber \\
	&=\frac{2}{\gamma^2+1}\tilde{F}_{ST}(x\gamma;\nu),
\end{align}
where $\tilde{F}_{ST}(\cdot;\nu)$ is the cdf of the $t$ distribution with $\nu$ degrees of freedom.

When $x\ge 0$ we obtain:
\begin{align}
	\label{eq:cdf2}
	F_{ST}(x;\gamma,\nu)&=\int_{-\infty}^xf_{ST}(t;\gamma,\nu)dt=\nonumber \\
	&=\frac{2}{\gamma^2+1}\tilde{F}_{ST}(0;\nu)+\frac{2}{\gamma+\frac{1}{\gamma}}\int_0^x \tilde{f}_{ST}\left(\frac{t}{\gamma};\nu\right)dt=\nonumber \\
	&=\frac{2}{\gamma^2+1}\frac{1}{2}+\frac{2}{\gamma+\frac{1}{\gamma}}\int_0^{x/\gamma} \tilde{f}_{ST}\left(u;\nu\right)\gamma du=\nonumber \\
	&=\frac{1}{\gamma^2+1}+\frac{2\gamma}{\gamma+\frac{1}{\gamma}}\left\{\tilde{F}_{ST}\left(\frac{x}{\gamma};\nu\right)-\tilde{F}_{ST}(0;\nu)\right\}=\nonumber \\
	&=\frac{1}{\gamma^2+1}+\frac{2\gamma}{\gamma+\frac{1}{\gamma}}\left\{\tilde{F}_{ST}\left(\frac{x}{\gamma};\nu\right)-\frac{1}{2}\right\}.
\end{align}

\subsection{Proof of Theorem \ref{teo:teo2}}

We consider the standardized case. We let $\alpha\in(0,1)$ and proceed by inverting (\ref{eq:stcdf}). We first need to compute $\alpha^*\stackrel{\text{def}}{=}F_{ST}(0;\gamma,\nu)$: if $\alpha<\alpha^*$, the resulting quantile will be smaller than 0, so that we invert (\ref{eq:cdf1}). We have:
\begin{align}
	&\frac{2}{\gamma^2+1}\tilde{F}_{ST}(x\gamma;\nu)=\alpha\Leftrightarrow \tilde{F}_{ST}(x\gamma;\nu)=\alpha\frac{\gamma^2+1}{2}\Leftrightarrow\nonumber \\
	&\Leftrightarrow x\gamma=\tilde{F}_{ST}^{-1}\left(\alpha\frac{\gamma^2+1}{2};\nu\right)\Leftrightarrow x=\frac{1}{\gamma}\tilde{F}_{ST}^{-1}\left(\alpha\frac{\gamma^2+1}{2};\nu\right)\nonumber.
\end{align}
If $\alpha\ge\alpha^*$, the resulting quantile will be larger than 0, so that we invert (\ref{eq:cdf2}). We have:
\begin{align}
	&\frac{1}{\gamma^2+1}+\frac{2\gamma}{\gamma+\frac{1}{\gamma}}\left\{\tilde{F}_{ST}\left(\frac{x}{\gamma};\nu\right)-\frac{1}{2}\right\}=\alpha\Leftrightarrow\nonumber \\
	&\Leftrightarrow\frac{2\gamma}{\gamma+\frac{1}{\gamma}}\left\{\tilde{F}_{ST}\left(\frac{x}{\gamma};\nu\right)-\frac{1}{2}\right\}=\alpha-\frac{1}{\gamma^2+1}\Leftrightarrow\nonumber \\
	&\Leftrightarrow\tilde{F}_{ST}\left(\frac{x}{\gamma};\nu\right)=\left(\alpha-\frac{1}{\gamma^2+1}\right)\left(\frac{\gamma^2+1}{2\gamma^2}\right)+\frac{1}{2}\Leftrightarrow\nonumber \\
	&\Leftrightarrow\tilde{F}_{ST}\left(\frac{x}{\gamma};\nu\right)=\frac{1}{2\gamma^2} \{\alpha(\gamma^2+1)-1\}+\frac{1}{2}\Leftrightarrow\nonumber \\
	&\Leftrightarrow x=\gamma\tilde{F}^{-1}_{ST}\left(\frac{1}{2\gamma^2} \{\alpha(\gamma^2+1)-1\}+\frac{1}{2};\nu\right).\nonumber
\end{align}

\subsection{Proof of Theorem \ref{teo:teo3}}

\begin{align}
	\text{E}(X_{ST})&=\frac{2}{\gamma+\frac{1}{\gamma}}\left\{\int_{-\infty}^0x\tilde{f}_{ST}(x\gamma;\nu)dx+\int_0^{\infty}x\tilde{f}_{ST}\left(\frac{x}{\gamma};\nu\right)dx\right\}=\nonumber \\
	&=\frac{2}{\gamma+\frac{1}{\gamma}}\left\{\int_{-\infty}^0\frac{1}{\gamma}\frac{u}{\gamma}\tilde{f}_{ST}(u;\nu)du+\int_0^{\infty}\gamma^2 u\tilde{f}_{ST}(u;\nu)du\right\}=\nonumber \\
	&=\frac{2}{\gamma^3+\gamma}\int_{-\infty}^0u\tilde{f}_{ST}(u;\nu)du+\frac{2\gamma^2}{\gamma+\frac{1}{\gamma}}\int_0^{\infty}u\tilde{f}_{ST}(u;\nu)du=\nonumber \\
	&=\frac{2}{\gamma^3+\gamma}\tilde{F}_{ST}(0;\nu) \underbrace{\int_{-\infty}^0u\frac{\tilde{f}_{ST}(u;\nu)}{\tilde{F}_{ST}(0;\nu)}du}_\text{$\text{E}(\tilde{X}_{ST,\nu}|\tilde{X}_{ST,\nu}<0)$}+\nonumber \\
	&+\frac{2\gamma^2}{\gamma+\frac{1}{\gamma}}[1-\tilde{F}_{ST}(0;\nu)]\underbrace{\int_0^{\infty} u\frac{\tilde{f}_{ST}(u;\nu)}{[1-\tilde{F}_{ST}(0;\nu)]}du}_\text{$\text{E}(\tilde{X}_{ST,\nu}|\tilde{X}_{ST,\nu}\ge0)$}.\nonumber
\end{align}
The conditional expectation $\text{CE}_{\nu}(a,b)\stackrel{\text{def}}{=}\text{E}(\tilde{X}_{ST,\nu}|a<\tilde{X}_{ST,\nu}<b)$ of the truncated $t_\nu$ distribution is given by \citep{kim08}:
\begin{equation}
	\label{eq:condexp}
	\xi_{\nu,a,b}\stackrel{\text{def}}{=}\text{E}(\tilde{X}_{ST,\nu}|a<\tilde{X}_{ST,\nu}<b)=
	\frac{\kappa\nu}{\nu-1}
	\left\{\left(1+\frac{a^2}{\nu}\right)^{-\frac{\nu-1}{2}}-\left(1+\frac{b^2}{\nu}\right)^{-\frac{\nu-1}{2}}\right\},
\end{equation}
with
$$
\kappa = \frac{\Gamma\left(\frac{\nu+1}{2}\right)}{\alpha_{0,j}\Gamma(\frac{\nu}{2})(\nu\pi)^{1/2}},\qquad
\alpha_{0,j} = \tilde{F}_{ST,\nu}(b) - \tilde{F}_{ST,\nu}(a).
$$
Noting that
$\text{E}(\tilde{X}_{ST,\nu}|\tilde{X}_{ST,\nu}<0)=-\text{E}(\tilde{X}_{ST,\nu}|\tilde{X}_{ST,\nu}\ge0)$, one gets:
$$
\text{E}(X_{ST})=-\xi_{\nu,0,+\infty}\frac{2}{\gamma^3+\gamma}\frac{1}{2}
+\xi_{\nu,0,+\infty}\frac{2\gamma^2}{\gamma+\frac{1}{\gamma}}\frac{1}{2}=
\xi_{\nu,0,+\infty}\left(\frac{\gamma^2}{\gamma+\frac{1}{\gamma}}-\frac{1}{\gamma^3+\gamma}\right).
$$
By rearranging the last expression, we have:
$$
\text{E}(X_{ST})=\xi_{\nu,0,+\infty}\left(\frac{\gamma^2-1}{\gamma}\right).
$$

\subsection{Proof of Theorem \ref{teo:teo4}}
Since $\text{var}(X_{ST})=\text{E}(X_{ST}^2)-\text{E}(X_{ST})^2$ and we have already computed $\text{E}(X_{ST})$, we focus on $\text{E}(X_{ST}^2)$. We have
\begin{align}
	\text{E}(X_{ST}^2)&=\frac{2}{\gamma+\frac{1}{\gamma}}\left\{\int_{-\infty}^0x^2\tilde{f}_{ST}(x\gamma;\nu)dx+\int_0^{\infty}x^2\tilde{f}_{ST}\left(\frac{x}{\gamma};\nu\right)dx\right\}=\nonumber \\
	&=\frac{2}{\gamma+\frac{1}{\gamma}}\left\{\int_{-\infty}^0\frac{u^2}{\gamma^3}\tilde{f}_{ST}(u;\nu)du+\int_0^{\infty}\gamma^3 u^2\tilde{f}_{ST}(u;\nu)du\right\}=\nonumber \\
	&=\frac{2}{\gamma^4+\gamma^3}\int_{-\infty}^0u^2\tilde{f}_{ST}(u;\nu)du+\frac{2\gamma^3}{\gamma+\frac{1}{\gamma}}\int_0^{\infty}u^2\tilde{f}_{ST}(u;\nu)du=\nonumber \\
	&=\frac{2}{\gamma^4+\gamma^3}\tilde{F}_{ST}(0;\nu) \underbrace{\int_{-\infty}^0u^2\frac{\tilde{f}_{ST}(u;\nu)}{\tilde{F}_{ST}(0;\nu)}du}_\text{$\text{E}(\tilde{X}^2_{ST,\nu}|\tilde{X}_{ST,\nu}<0)$}+\nonumber \\
	&+\frac{2\gamma^3}{\gamma+\frac{1}{\gamma}}[1-\tilde{F}_{ST}(0;\nu)]\underbrace{\int_0^{\infty} u^2\frac{\tilde{f}_{ST}(u;\nu)}{[1-\tilde{F}_{ST}(0;\nu)]}du}_\text{$\text{E}(\tilde{X}^2_{ST,\nu}|\tilde{X}_{ST,\nu}\ge0)$}.\nonumber
\end{align}
The conditional expectation $\text{E}(\tilde{X}^2_{ST,\nu}|a<\tilde{X}_{ST,\nu}<b)$ is given by \citep{kim08}:
\begin{align}
	\label{eq:ce2}
	\xi^2_{\nu,a,b}&\stackrel{\text{def}}{=}\text{E}(\tilde{X}^2_{ST,\nu}|a<\tilde{X}_{ST,\nu}<b)=\nonumber \\
	&=\frac{\nu-1}{\tau_{1,j}} \left\{\frac{\tilde{F}_{ST}(b\tau_1^{1/2},\nu-2)-\tilde{F}_{ST}(a\tau_1^{1/2},\nu-2)}{\tilde{F}_{ST}(b,\nu)-\tilde{F}_{ST}(a,\nu)}\right\} - \nu,\nonumber
\end{align}
where $\tau_1=(\nu-2)/\nu$.

Noting that
$\text{E}(\tilde{X}^2_{ST,\nu}|\tilde{X}_{ST,\nu}<0)=\text{E}(\tilde{X}^2_{ST,\nu}|\tilde{X}_{ST,\nu}>0)$, one gets:
$$
\text{E}(X^2_{ST})=\frac{2}{\gamma^4+\gamma^3}\frac{1}{2} \xi^2_{\nu,0,+\infty}
+\frac{2\gamma^3}{\gamma+\frac{1}{\gamma}}\frac{1}{2}\xi^2_{\nu,0,+\infty}=\xi^2_{\nu,0,+\infty}\left(\frac{1}{\gamma^4+\gamma^3}+\frac{\gamma^3}{\gamma+\frac{1}{\gamma}}\right).
$$
Finally, we have:
\begin{align}
	\text{var}(X_{ST})&=\xi^2_{\nu,0,+\infty}\left(\frac{1}{\gamma^4+\gamma^3}+\frac{\gamma^3}{\gamma+\frac{1}{\gamma}}\right)-[\text{E}(X_{ST})]^2\nonumber \\
	&=\xi^2_{\nu,0,+\infty}\frac{\gamma^6+1}{(\gamma^2+1)\gamma^2}-\left(\xi_{\nu,0,+\infty}\frac{\gamma^2-1}{\gamma}\right)^2=\nonumber \\
	&=\xi^2_{\nu,0,+\infty}\left\{\frac{\gamma^6+1}{(\gamma^2+1)\gamma^2}-\left(\frac{\gamma^2-1}{\gamma}\right)^2\right\}.\nonumber 
\end{align}

\end{appendix}

\end{document}